\documentclass[twocolumn]{aastex63}
\usepackage{graphicx}
\usepackage{amsmath}
\usepackage{multirow}
\newcolumntype{C}{>{\centering\arraybackslash}X}
\newcommand{\xj}[1]{ { {#1}} }

\newcommand{\Oabund}{0.41$^{+0.20}_{-0.30}$}
\newcommand{\Siabund}{0.74$^{+0.36}_{-0.53}$}
\defcitealias{Richard2021}{R21}

\received{xxx}
\revised{xxx}
\accepted{xxx}
\submitjournal{ApJL}

\shorttitle{Metal-Enriched Neutral Gas around Low-mass Galaxies at $z=4$}
\shortauthors{X.Lin et al.}
\graphicspath{{./}{figures/}}

\begin{document}

\title{Metal-Enriched Neutral Gas Reservoir around a Strongly-lensed, Low-mass Galaxy at $z=4$ Identified by JWST/NIRISS and VLT/MUSE}

\author[0000-0001-6052-4234]{Xiaojing Lin}
\affiliation{Department of Astronomy, Tsinghua University, Beijing 100084, China\email{linxj21@mails.tsinghua.edu.cn, zcai@mail.tsinghua.edu.cn}}

\author[0000-0001-8467-6478]{Zheng Cai}
\affiliation{Department of Astronomy, Tsinghua University, Beijing 100084, China\email{linxj21@mails.tsinghua.edu.cn, zcai@mail.tsinghua.edu.cn}}

\author[0000-0002-3983-6484]{Siwei Zou}
\affiliation{Department of Astronomy, Tsinghua University, Beijing 100084, China\email{linxj21@mails.tsinghua.edu.cn, zcai@mail.tsinghua.edu.cn}}

\author[0000-0001-5951-459X]{Zihao Li}
\affiliation{Department of Astronomy, Tsinghua University, Beijing 100084, China\email{linxj21@mails.tsinghua.edu.cn, zcai@mail.tsinghua.edu.cn}}

\author[0000-0002-2178-5471]{Zuyi Chen}
\affil{Steward Observatory, University of Arizona, 933 N Cherry Ave, Tucson, AZ 85721, USA}

\author[0000-0002-1620-0897]{Fuyan Bian}
\affiliation{European Southern Observatory, Alonso de C\'{o}rdova 3107, Casilla 19001, Vitacura, Santiago 19, Chile\email{Fuyan.Bian@eso.org}}

\author[0000-0002-4622-6617]{Fengwu Sun}
\affiliation{Steward Observatory, University of Arizona, 933 N Cherry Ave, Tucson, AZ 85721, USA}

\author{Yiping Shu} 
\affiliation{Purple Mountain Observatory, Chinese Academy of Sciences, 10 Yuan Hua Road, Nanjing, Jiangsu 210023, China}

\author[0000-0003-0111-8249]{Yunjing Wu}
\affiliation{Department of Astronomy, Tsinghua University, Beijing 100084, China\email{linxj21@mails.tsinghua.edu.cn, zcai@mail.tsinghua.edu.cn}}
\affil{Steward Observatory, University of Arizona, 933 N Cherry Ave, Tucson, AZ 85721, USA}

\author[0000-0001-6251-649X]{Mingyu Li}
\affiliation{Department of Astronomy, Tsinghua University, Beijing 100084, China\email{linxj21@mails.tsinghua.edu.cn, zcai@mail.tsinghua.edu.cn}}

\author[0000-0002-1815-4839]{Jianan Li} 
\affiliation{Department of Astronomy, Tsinghua University, Beijing 100084, China\email{linxj21@mails.tsinghua.edu.cn, zcai@mail.tsinghua.edu.cn}}

\author[0000-0003-3310-0131]{Xiaohui Fan}
\affil{Steward Observatory, University of Arizona, 933 N Cherry Ave, Tucson, AZ 85721, USA}

\author[0000-0002-7738-6875]{J. Xavier Prochaska}
\affiliation{Department of Astronomy and Astrophysics, UCO, Lick Observatory, University of California
1156 High Street, Santa Cruz, CA 95064, USA}
\affiliation{Kavli IPMU, the University of Tokyo (WPI), Kashiwa 277-8583, Japan}

\author[0000-0001-7144-7182]{Daniel Schaerer}\affiliation{Observatoire de Geneve, Universite de Geneve, Chemin Pegasi 51, CH-1290 Versoix, Switzerland}

\author[0000-0003-3458-2275]{Stephane Charlot}\affiliation{Sorbonne Universit\'e, CNRS, UMR7095, Institut d'Astrophysique de Paris, F-75014, Paris, France}

\author[0000-0002-8726-7685]{Daniel Espada}
\affiliation{Departamento de F\'{i}sica Te\'{o}rica y del Cosmos, Campus de Fuentenueva, Edificio Mecenas, Facultad de Ciencias, Universidad de Granada, E-18071, Granada, Spain}
\affiliation{Instituto Carlos I de F\'{i}sica Te\'{o}rica y Computacional, Facultad de Ciencias, E-18071, Granada, Spain}

% \author[0000-0002-5057-135X]{Eros Vanzella}
% \affiliation{INAF – OAS, Osservatorio di Astrofisica e Scienza dello Spazio di Bologna, via Gobetti 93/3, I-40129 Bologna, Italy}

\author{Miroslava Dessauges-Zavadsky}
\affiliation{Department of Astronomy, University of Geneva, Chemin Pegasi 51, 1290 Versoix, Switzerland}

\author[0000-0003-1344-9475]{Eiichi Egami}
\affiliation{Steward Observatory, University of Arizona, 933 N Cherry Ave, Tucson, AZ 85721, USA}

\author{Daniel Stark}
\affil{Steward Observatory, University of Arizona, 933 N Cherry Ave, Tucson, AZ 85721, USA}

\author{Kirsten K. Knudsen}
\affiliation{Institutionen för rymd-, geo- och miljövetenskap, Department of Space, Earth and Environment
Chalmers tekniska högskola, Chalmers University of Technology}

% \author[0000-0003-3926-1411]{Jorge Gonz\'alez-L\'opez}
% \affil{Las Campanas Observatory, Carnegie Institution of Washington, Casilla 601, La Serena, Chile}
% \affil{N\'ucleo de Astronom\'ia de la Facultad de Ingenier\'ia y Ciencias, Universidad Diego Portales, Av. Ej\'ercito Libertador 441, Santiago, Chile}

\author{Gustavo Bruzual}\affiliation{Institute of Radio Astronomy and Astrophysics, National Autonomous University of Mexico, San José de la Huerta
58089 Morelia, Michoacán, México}

\author[0000-0002-7636-0534]{Jacopo Chevallard}
\affiliation{Department of Physics, University of Oxford, Denys Wilkinson Building, Keble Road, Oxford OX1 3RH, UK}

\begin{abstract}

Direct observations of low-mass, low-metallicity galaxies at $z\gtrsim4$ provide an indispensable opportunity 
for detailed inspection of the ionization radiation, gas flow, and metal enrichment in sources similar to 
those that reionized the Universe. 
Combining the James Webb Space Telescope (JWST), VLT/MUSE, and ALMA, we present detailed observations of a strongly lensed,  low-mass ($\approx 10^{7.6}$ ${\rm M}_\odot$) galaxy at $z=3.98$ (also see \citealt{Vanzella2022}). We identify strong narrow nebular emission, including  \ion{C}{4}~$\lambda\lambda1548,1550$, \ion{He}{2}~$\lambda1640$, \ion{O}{3}]~$\lambda\lambda1661,1666$, [\ion{Ne}{3}]~$\lambda3868$, [\ion{O}{2}]~$\lambda3727$, and Balmer series of Hydrogen from this galaxy, 
indicating a metal-poor \ion{H}{2} region ($\lesssim 0.12\ {\rm Z}_\odot$) powered by massive stars. 
Further, 
we detect a metal-enriched  
 damped Ly$\alpha$ system (DLA) associated with the galaxy 
 with the \ion{H}{1} column density of $N_{\rm{HI}}\approx 10^{21.8}$ cm$^{-2}$. 
 The metallicity of the associated DLA may reach the  
super solar metallicity (${\gtrsim Z}_\odot$). 
Moreover, thanks to JWST and gravitational lensing, we present the resolved UV slope ($\beta$) map at the 
spatial resolution of $\approx 100$ pc at $z=4$, with steep UV slopes  
reaching $\beta \approx -2.5$ around 
three star-forming clumps. Combining with low-redshift analogs, our observations suggest that low-mass, 
low-metallicity galaxies, which  
dominate reionization, 
could be surrounded by a high 
covering fraction of the metal-enriched, neutral-gaseous clouds. This implies that the metal enrichment of 
low-mass galaxies is highly efficient, and further support that 
in low-mass galaxies, only a small fraction of ionizing radiation can escape 
through the interstellar or circumgalactic channels with low column-density neutral gas. 
 
\end{abstract}

\keywords{DLA --- host galaxy --- lensing --- high-redshift --- reionization}

\section{Introduction}

%Low-mass, low-metallicity galaxies have great cosmological significance \citep[e.g.,][]{Berg2018}.
It is now %generally 
believed that low-mass galaxies significantly 
contribute the major fraction of the  %required 
ionization photons in the epoch 
of reionization \citep[e.g.,][]{Atek2015,Bouwens2016}. %Nevertheless, direct observations of such galaxies 
%in the faint-end luminosity function are difficult. 
Observations of faint galaxies at high-redshift, though very difficult, are required, especially in the era of James Webb Space Telescope (JWST), to directly constrain the ionization 
radiation, gas environment, galaxy growth, and metal enrichment of the surrounding intergalactic medium (IGM). 
% Although the direct observations of faint galaxies at high-redshift are difficult, 
% such observations 
% are required, especially in the era of JWST, to directly constrain the ionization 
% radiation, gas environment, galaxy growth, and metal enrichment of the surrounding intergalactic medium (IGM). 

% Further, low-mass, low-metallicity galaxies have great cosmological significance \citep[e.g.,][]{Berg2018}. %\citep{Brammer2012,Bian2017,Berg2018,Erb2019}. 
% Extensive efforts have been conducted to study the intrinsic continuum and emission lines, aimed to study whether 
% and how the ionization radiation can escape into the intergalactic medium (IGM) \citep[e.g.,][]{Vasei2016}.
% %\citep{Vasei2016,Oesch2018,Bouwens2022}. 

%%%%%----------------------------PAY ATTENTION TO HERE-------------------------------------
%%%%%% consider to use \cite[e.g.,][]{one reference} in above paragraph for not very crucial sentences
%%%%%% Reference limit to journal could be 50 (Arxiv version can cite a bit more). 
%%%%%%%------------------------------------------------------------------------------------

%Massive stars in the low mass, low metallicity galaxies can yield significant Lyman continuum radiation of the ionizing photons, 
%with a steep UV slope \cite[e.g.,][]{Bouwens2010}. 
Extensive efforts have been conducted to study the intrinsic 
ultra-violet (UV) continuum and the ionization of low-mass galaxies.  \xj{The ultraviolet (UV) continuum slope $\beta$ ($f_\lambda\propto\lambda^\beta$) reveals crucial properties of high-redshift galaxies. It is believed that young stellar population and intense star formation in early galaxies lead to very blue UV slopes \citep[e.g.][]{Schaerer2003}, even reaching $-3$ with extremely efficient ionizing agents with minimal dust reddening \citep[e.g.][]{Topping2022}. The high spatial resolution of JWST enables studies on spatial variations of $\beta$, mapping the distribution of stellar populations and star forming regions within early galaxies, often with clumpy structures, on sub-kiloparsec scales \citep{Chen2022}.}

A great many studies aim to understand how ionization radiation escapes into the IGM \citep[e.g.,][]{Ramambason2020}. %,Saldana-Lopez2022{\bf add references here} 
 Before UV photons reach the IGM, how much radiation can escape 
 from galaxies depends on: 
(1) the intrinsic intensity and the slope of UV spectra; and 
(2) the distribution of optically-thick neutral hydrogen (\ion{H}{1}) gas in the 
interstellar and circumgalactic medium (CGM) \cite[e.g.,][]{Chisholm2018}. %,Gazagnes2020}. 
It is therefore crucial to study %properties of 
both UV continua and gas properties around 
low-mass galaxies to quantify the \ion{H}{1} (Ly$\alpha$) 
optical depth in the CGM for better understanding the 
reionization process.
%in order to further  %\cite[e.g.,][]{Alavi2020,Izotov2021}. 

%The observations of th 
The %\ion{H}{1} 
existence of the neutral gas in and around %high-$z$ 
low-mass star forming galaxies 
has been suggested mainly through 
observations of Ly$\alpha$ emission \citep[e.g.,][]{Stark2017,Gronke2017}.  
%and 
%{absorption \citep[e.g.,][]{not14,kro17}.}
%Observations of 
Low-mass galaxies often have strong Ly$\alpha$ and other ionized nebular emission \citep[e.g.][]{Erb2010,Stark2014}. 
The presence of optically-thick, \ion{H}{1} ``shell"  makes the Ly$\alpha$ 
emission %lines show the 
showing the double-peak structure \citep[e.g.,][]{Yang2017}, 
with the expanding outflow further suppressing blue peaks of Ly$\alpha$ emission \citep{Yang2014}. %{The damped Ly$\alpha$ systems (DLA, \citealt{wolfe05}) having N~({H~\sc i}) $>$ 10$^{20.3}$ cm$^{-2}$ seen in the absorption systems towards bright background sources are detected around the Lyman-break galaxies and Lyman-$\alpha$ emitters.}  

Nevertheless, %the modeling of Ly$\alpha$ emission line only produces limited information.
 {only through emission, the nature of the interplay between the neutral gas and 
 galaxy cannot be completely understood.}
Detailed gas properties, including the gas metallicity, geometry and density 
need to be constrained. 
%However, for the low-mass galaxy at high-redshift, %direct observations of their rest-frame UV continuum, nebular emission lines of ionized gas, and the \ion{H}{1} neutral gas remain to be highly limited. 
%the faintness of the galaxy continuum makes it extremely hard to study gas properties through absorption. 
%The limited wavelength coverage also 
%makes it difficult to have complete information of ionized gas. 
%The limited spatial resolution further hinders the understanding the spatial variations of 
%the anisotropic radiation of ionizing photons \citep[e.g.,][]{Gazagnes2020}. 
The strong lensing, plus the deep integration of JWST and VLT/MUSE provide us with 
an excellent opportunity to conduct direct and detailed observations of a low-mass 
galaxy at $z=4$, through both emission and absorption. Here, we present a detailed analysis of a strongly magnified arclet 
galaxy in the Abell2744 field, combined with optical, infrared and millimeter observations. 
{ The system is lensed into three multiple images, and the arclet is composed of two of them 
lying on both sides of the critical line, with three bright compact clumps (knots in the arc) mirrored on each side 
\citep{Mahler2018,Bergamini2022}. We name this galaxy ``A2744-arc3". % in this paper.  
It was also reported in \citet{Vanzella2022}. 
Through spectral-energy-distribution (SED) fitting, they found that 
these clumps are gravitationally-bound young massive star clusters  with bursty star formation.
% with ages of a few tens of Myr, 
% stellar masses of $(0.7-4.0)\times 10^6 {\rm M}_\odot$, optical/ultraviolet effective radii of $3 < R_{\rm eff} < 20$ pc 
% and bursty sSFR of 10 Gyr$^{-1}$. 
}
%  The VLT/MUSE and JWST/NIRISS yield the wide imaging and spectral 
% coverage from 0.5 -- 2.2 $\mu$m of this galaxy. 
% The spectral-energy-distribution (SED) fitting allows us to understand the 
% properties of the star clumps (e.g., age, stellar mass) (also see \citealt{Vanzella2022}).

{In this paper, we probe detailed properties of the interstellar and circumgalactic medium (CGM) of A2744-arc3. With VLT/MUSE and JWST, we provide  
%by probing the spectral features of A2744-arc3. 
detailed analysis of both nebular emission and absorption lines. %of this galaxy are 
%identified and analysed. 
% Combining this galaxy with low-redshift 
% analogs, we aim to summarize a global picture in low-mass galaxies. 
}
%{better delete the combining..sentence?}
%go deep into the spectral features of A2744-arc3. } 
Copious emission lines, including the detection of  
\ion{C}{4}~$\lambda\lambda1548,1550$, \ion{He}{2}~$\lambda1640$, \ion{O}{3}]~$\lambda\lambda1661,1666$, [\ion{Ne}{3}]~$\lambda3868$, [\ion{O}{2}]~$\lambda3727$, and Balmer series of Hydrogen, 
% {\bf double check, consistent with abstract},  
allow us to quantify the low metallicity of the ionized medium \citep[e.g.,][]{Smit2017,Witstok2021}. 
Further, the high signal-to-noise (S/N) observations of absorption 
enable us, for the first time, to 
examine the detailed properties of the neutral gas reservoir surrounding a low-mass 
galaxy at $z\approx4$. The excellent de-lensed spatial resolution down to $\approx100$ pc in rest-frame UV observations %allow us 
 allow the examination of the spatial variations of the UV radiation. All these 
information yield unparalleled understanding of low-mass, young galaxies at high-redshift  
using A2744-arc3 as a textbook example. 

This paper is organized as follows. In Section \ref{sec:data}, we describe the observations and data reduction, including JWST/NIRISS imaging and spectroscopy, HST imaging, VLT/MUSE and ALMA continuum observations; In Section \ref{sec:result}, we provide details of photometry, SED fitting, and the lensing reconstruction in source-plane. We also show the spatially resolved UV slope map. In Section \ref{sec:dla}, we present further analysis on the rest-frame UV and optical spectrum of A2744-arc3, including the analysis of both emission and absorption features. Discussions are conducted in Section \ref{sec:discussion}. We assume a flat cosmological model with $\rm{\Omega_{M} = 0.3, \Omega_{\Lambda} = 0.7}$ and $\rm{H_0 = 70\ km\ s^{-1}\ Mpc^{-1}}$.  The \texttt{Glafic} lensing model \citep{Oguri2010,Kawamata2016} is adopted.

\section{Data}\label{sec:data}
\subsection{JWST/NIRISS imaging \& grism observations}\label{sec:niriss data}
NIRISS observations of A2744 were taken by the
GLASS JWST Early Release Science Program \citep{Treu2022}, devoting about $2834.504\,\mathrm{s} \simeq 0.79\,\mathrm{hours}$ for direct imaging and $2\times 5196.5 {\rm s}\simeq 2.9{\rm hr}$ for GR150 row and column (R+C) grism in F115W, F150W, and F200W. {Especially the NIRISS/F200W data, spanning from 1.7$\mu{\rm m}$ to 2.2 $\mu{\rm m}$, allows us to better improve constraints on its stellar mass.} The data were reduced in the same way as shown in \citet{Wu2022}. Here we summarize the procedure as follows: the data were reduced using the standard JWST pipeline\footnote{\href{https://github.com/spacetelescope/jwst}{https://github.com/spacetelescope/jwst}} v1.6.2 with calibration reference files ``\texttt{jwst\_0944.pmap}''; 
1/f noise (see \citealt{Schlawin20}) was modeled and removed using the code \texttt{tshirt/roeba}\footnote{\href{https://github.com/eas342/tshirt}{https://github.com/eas342/tshirt}}, and the ``snowball'' artifacts from cosmic rays \citep{Rigby22} were identified and masked; the world coordinate system of mosaicked images were registered using the Pan-STARRS1 catalog \citep{panstarrsdr1}; the pixel scale of final mosaicked images was resampled to 0.03\arcsec\ with \texttt{pixfrac}\,$=$\,0.8. We reduced NIRISS grism data using \textsc{Grizli} \footnote{\url{https://github.com/gbrammer/grizli/}}, and the 2D grism spectra were drizzled with a pixel scale 0.065\arcsec.

To optimally extract 1D spectra of A2744-arc3, we first split the entire arclet into six segments, each with one clump. We extracted their 1D spectra using 5-th Chebyshev polynomials. We summed over five 1D spectra of the six segments to get the final spectra. One of them was excluded since it is severely contaminated by zero-th and first order images from other sources. 

\subsection{Hubble Frontier Fields images}

The A2744 cluster was observed as part of the Hubble Space Telescope Frontier Fields program\footnote{\url{https://archive.stsci.edu/prepds/frontier/}} (HFF; ID: 13495, PI: J. Lotz) 
between 2013 October 25 and 2014 July 1 for 164.5 orbits in total, in three filters with the Advanced Camera for Surveys (ACS; F435W,
F606W, F814W) and four with the Wide Field Camera 3
(WFC3; F105W, F125W, F140W and F160W).  The high-level science product released by Space Telescope
Science Institute (STScI) with a pixel size of 30 mas is used in this study. 

\subsection{VLT/MUSE observation}

Multi Unit Spectroscopic Explorer (MUSE) observations toward A2744 were taken by the GTO Program 094.A-0115 (PI: Richard) between 2014 September and 2015 October, with a 2 arcmin$\times$2 arcmin designed mosaic of pointings centred at $\alpha$= 00h14m20.952s and $\delta$= -30$^\circ$23$^\prime$53.88$^{\prime\prime}$ \citep{Mahler2018}. Each pointing was observed for a total of 3.5, 4, 4 and 5h in addition to 2h at the cluster center. The publicly available data cube\footnote{\url{http://muse-vlt.eu/science/a2744/}}  is adopted in this work.

\subsection{ALMA continuum observations}

The millimeter continuum observations are obtained from the ALMA Lensing Cluster Survey (ALCS).
A2744 is observed by ALMA band 6 (Program 2018.1.00035.L, PI: Kohno; 2013.1.00999.S, PI: Bauer) which covers a frequency range with two tunings of $250.0-257.5$ GHz and $265.0-272.5$ GHz, corresponding to wavelengths of $\sim 1.15 $ mm \citep{Sun2022}.
The $1\sigma$ continuum sensitivity or say rms is 45\,$\mu$Jy\,beam$^{-1}$.
% The effective time on A2744-arc3 is about 8 minutes. %{\bf ZC: indeed a bit short..}. 
The data were reduced using the standard pipeline with the CASA \citep{McMullin2007} with natural weighting, and will be described in details in S. Fujimoto et al., in prep. This leads to a synthesized beam size (FWHM) of $\sim 1$\arcsec, corresponding to $\sim6.9$ kpc at $z\sim 4$.

\section{Result and Analysis}\label{sec:result}

%  The system is lensed into three multiple images, and the arclet is composed of two of them 
% lying on both sides of the critical line, with three bright compact clumps (knots in the arc) mirrored on each side 
% \citep{Mahler2018,Bergamini2022}. 

% A2744-arc3 is a strongly magnified arclet centered at $\alpha=00$h14m21.38s and $\delta=-30^\circ$23$^\prime$37.8$^{\prime\prime}$.  The system is lensed
% into three multiple images \citep{Mahler2018,Bergamini2022}, two of which 
% reside on both sides of the critical line, consisting of the entire arclet we observed with three bright compact clumps mirrored on each side. As introduced in \cite{Vanzella2022},  the three clumps are young (a few tens of Myr), bursty star-forming (sSFR $\sim $ 10 Gyr$^{-1}$) and compact with stellar mass surface densities resembling those of local globular clusters. The magnification varies from 20 to $>$1000 across the arc and 20-200 for the clumps, based on \texttt{Glafic} lensing model. 

\subsection{Photometry and SED fitting}\label{sec:photometry}

We use a box aperture along A2744-arc3 to carry out the photometry on seven broad-band HST observations (ACS/F435W, F606W, F814W and WFC3/F105W,  F125W, F140W, and F160W) and three JWST NIRISS observations (NIRISS/F115W, F150W, and F200W). For each band, the length and width of the box aperture are designed to match the 2$\sigma$ contours of the arc, where $\sigma$ is determined by the local sky background. A sersic modelling is performed to the nearest source in the north to exclude potential light contamination. Magnification and multi-image effects are corrected to determine the delensed flux. {The median magnification along the arc is about 45 according to the \texttt{Glafic} model. We repeat the photometry and SED analysis using \citet{Richard2021} model, as described in Appendix \ref{sec:lensing model systematics}, and yield consistent results.}

\begin{figure*}
    \includegraphics[width=\textwidth]{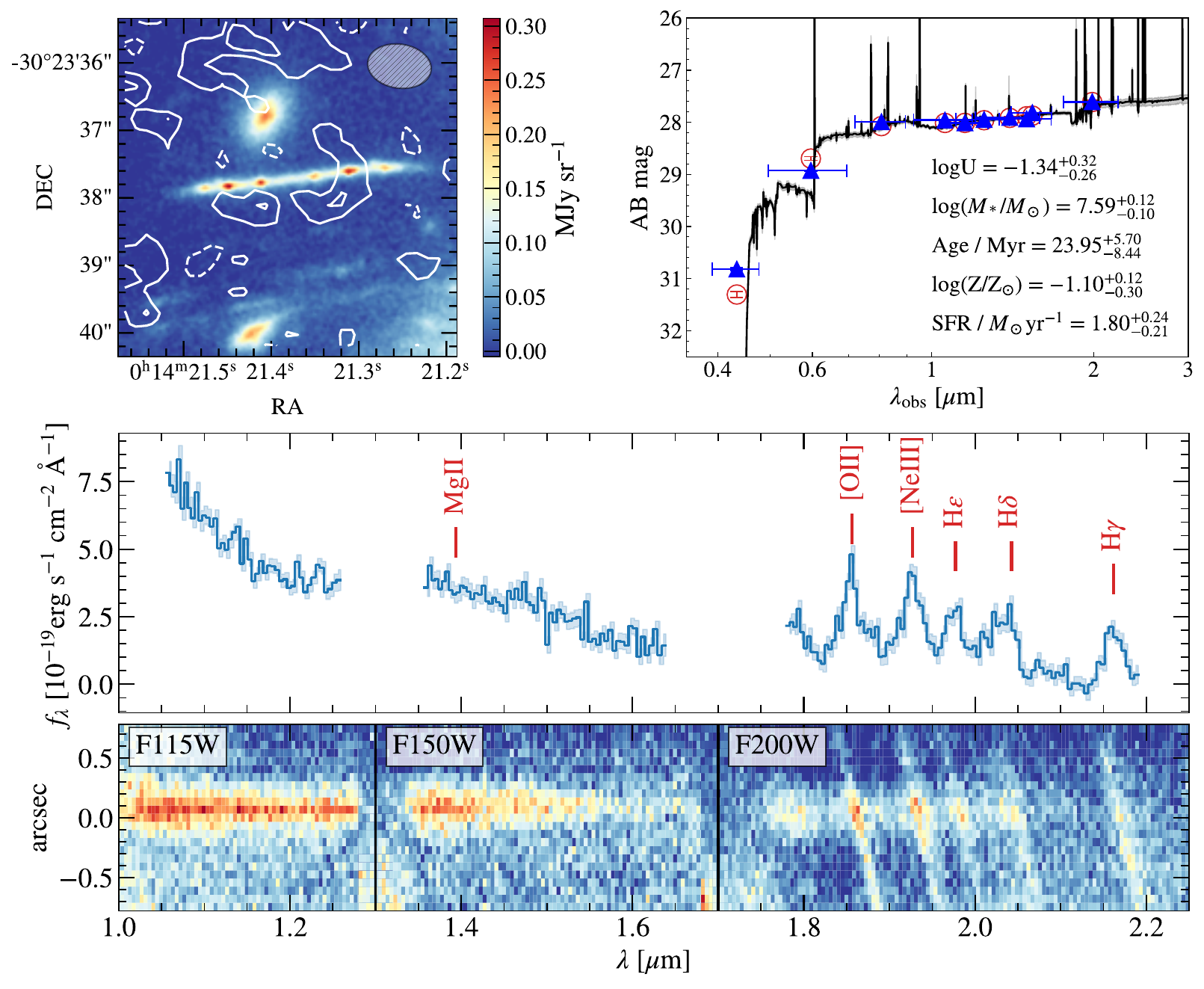}
    \caption{
    % {\bf please confirm the unit is $\mu$Jy rather than MJy.} 
    {\bf Top left}: JWST NIRISS F150W surface brightness image of A2744-arc3 with overlaid 
    continuum contours of ALMA band 6. The contours levels are -2 (white dashed), 1 and 2$\sigma$ (white solid) with $1\sigma=44.5\mu$Jy/beam, 
    showing no clear detection at a rest-frame wavelength of 0.23mm.  {\bf Top right}: Photometry and SED fitting for A2744-arc3, with derived physical parameters. Blue triangles denote the measurement and red circles the best-fit result. {\bf Bottom}: Optimally extracted 1D spectra (top panel) and joint 2D spectra covered by NIRISS three filters (F115W, F150W and F200W). In the top panel, shaded blue regions indicate flux uncertainties, and vertical red lines label common emission lines, among which \ion{Mg}{2} is not detected. The 2D spectra of the bottom panel are obtained by summing 2D spectra of the five segments as described in Section \ref{sec:niriss data}}
    \label{fig:jwst}
\end{figure*}

We perform SED modeling with the Bayesian code \texttt{BEAGLE} \citep{Chevallard2106}. We assume a constant star formation history, the dust extinction curve of the Small Magellanic Cloud with the optical depth in the V band varying in the range 0 -- 0.5, and the \cite{Chabrier2003} initial mass function (IMF) with an upper limit of 300${\rm M}_\odot$. The metallicity of the interstellar medium is kept the same as that of the stellar populations. A flat prior is applied for metallicity, $0.01<{ Z}/{\rm Z}_\odot<0.2$, and ionization parameter (the dimensionless ratio of the number density of H-ionizing photons to that of hydrogen) in log-space, $-3<\log{U}<-1$, for nebular emission \citep{Gutkin2016}. 
{These %low-$Z$, high-$U$ 
priors are chosen on the basis of detections of the UV emission lines (see Section \ref{sec:emission} for more details)}. 
A JWST NIRISS/F150W view of A2744-arc3, the SED fitting results, and the NIRISS grism 2D and 1D spectra are shown in Figure \ref{fig:jwst}. It implies that A2744-arc3 is a young, low-mass galaxy with massive 
star formation, {accompanied by blue UV slopes, strong nebular emission lines and a faint intrinsic UV luminosity}.  {The absence of a Balmer break in the NIRISS spectrum further confirms a %relatively 
young stellar age.}
% A more detailed discussion on its metallicity associated with nebular emission lines is conducted in Section \ref{sec:emission}.  
The continuum of ALMA band 6 at rest-frame 0.23\,mm shows no clear detection, which indicates a 3$\sigma$ upper limit of far-infrared luminosity surface density $<2.2\times 10^{11}{\rm L}_\odot/{\rm arcsec^2}$ assuming a dust temperature $T=35{\rm K}$ and dust emissivity spectral index $b=1.6$. This corresponds to an upper limit of star formation rate surface density $\approx 9\times 10^{-7} {\rm M}_\odot{\rm yr^{-1} pc^{-2}}$ \citep{Murphy2011}. %{The non-detection in submilimeters implies lack of dust in A2744-arc3.} 
{and an upper limit of dust mass surface density $\approx 1.6 {\rm M}_\odot {\rm pc}^{-2}$ \citep{daCunha2013}.}

{In \citet{Vanzella2022}, SED analysis was performed on each mirrored clump using images from NIRISS/F115W, F150W, F200W, and HST/WFC3 F105W, F125W, F140W, F160W, along with the ground-based VLT/HAWKI Ks-band. 
{Their derived star formation rate (SFR) is $1.47^{+0.85}_{
-0.25} /{\rm M}_\odot {\rm yr}^{-1}$, total stellar mass  of $1.6
^{+0.4}_{-1.3}\times 10^{8} {\rm M}_\odot$ (Table 1, 3c), 
and age of $126^{+33}_{-108}$ Myr. Our derived properties are marginally consistent with theirs within $1\sigma$.} They found that these nucleated star-forming regions have optical/UV effective radii spanning 3--20 pc, bursty star formation activities ($\sim 10 {\rm Gyr}^{-1}$) and UV slopes of $\beta \simeq -1.9$, and  stellar mass surface densities  resembling those of local %young %massive star clusters and 
globular clusters. %suggesting that these compact clumps might be globular cluster precursors. 
Our conclusion is consistent with theirs. 
%, further 
%confirming that A2744-arc3 hosts young and low-mass stellar populations.  
%with intense star formation.  
Note that the upper limit of IR-based SFR density is smaller than that obtained from UV 
%in UV-estimated SFR density in 
($\gtrsim 10^{-5} {\rm M}_\odot{\rm yr^{-1} pc^{-2}}$ for clumps in \citet{Vanzella2022}), and 
this may be due to a low dust-obscured fraction of SFR in this low-mass galaxy 
with negligible dust content. % the lack of FIR emission originating from dust heating.
}

% \subsection{Source Plane Reconstruction} \label{sec:reconstruction}

% We recover the intrinsic source morphology by ray-tracing back the image-plane pixels on a regular source plane grid based on the corresponding deflection field \citep{Patricio2016}.

% Limited by the lensing model, we can hardly reconstruct the three clump pairs  onto the source plane simultaneously, where each pair should have perfectly overlapped. Among all publicly released lensing models, the \texttt{Glafic} model presents the smallest offsets between source-plane positions of these clumps pairs, yielding $0.010''$, 0.017$^{\prime\prime}$, and 0.033$^{\prime\prime}$ from inside to outside respectively. To avoid model-dependent systematics introduced by the non-overlap, the reconstruction is only performed on the west side of the arclet that shows higher S/Ns for the three separated clumps.

\subsection{Source Plane Reconstruction and Spatially-resolved UV Slope}\label{sec:reconstruction}

% The ultraviolet (UV) continuum slope $\beta$ ($f_\lambda \propto \lambda^{\beta}$) works a prominent metric for properties of high-redshift galaxies. Blue UV continuum slopes, even lower than -3, are predicted 
% for young and low metallicity stellar populations, with less attenuation from dust \citep{Topping2022}.  
To explore the spatially-resolved UV slope variation, we construct UV slope $\beta$ maps in both 
the image plane and source plane using HST and JWST broad-band images.  We recover the intrinsic source morphology by ray-tracing back the image-plane pixels on a regular source plane grid based on the corresponding deflection field \citep{Patricio2016}. {More details about the selection of lensing models and the reconstruction are presented in Appendix \ref{sec:lensing model systematics}.}

\begin{figure*}[htbp]
    \centering
    \includegraphics[width=\textwidth]{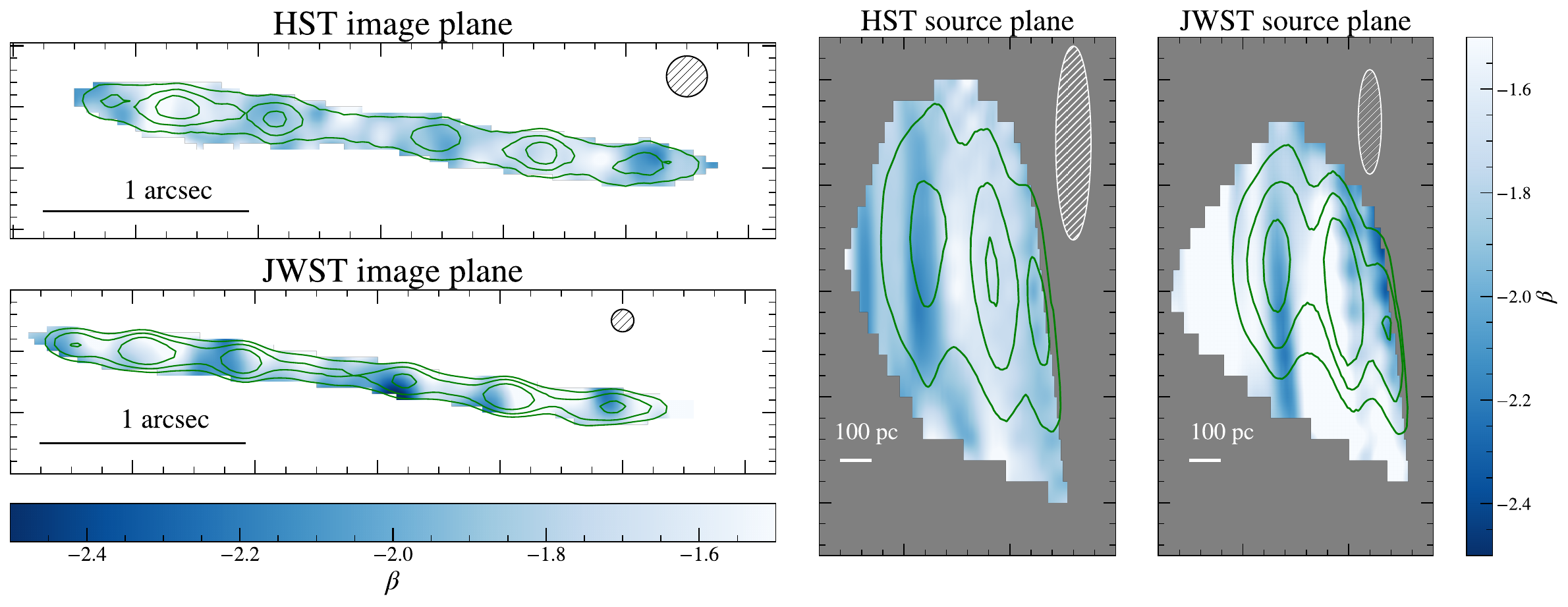}
    \caption{
    % {\bf Enlarge the title: ``HST image plane, HST source plan and JWST source plane" }
    The UV slope $\beta$ map in the image plane and source plane measured from HST and JWST broadband images. The HST UV slope maps are measured from WFC3/F105W, F125W, F140W and F160W images (2125-3091\,\AA\ in the rest frame), while the JWST UV slope maps are constructed only using NIRISS/F115W and F150W images (2315-3015\,\AA\ in the rest frame), to avoid emission lines captured by the NIRISS/F200W filter. The overlaid green contours in HST image/source plane denote the lensed/de-lensed WFC3/F160W surface brightness levels of 0.05,0.075,0.1 MJy sr$^{-1}$,
    %{\bf should be $\mu$Jy sr$^{-1}$? \  \AA\ should not be italic throughout the paper} 
    and the contours in JWST image/source plane show NIRISS/F150W lensed/de-lensed surface brightness levels of 0.075,0.1,0.135 MJy sr$^{-1}$ respectively. The ellipses on the upper right corners correspond to the FWHM of image/source plane point spread function (PSF). Note that source-plane 
    PSF varies among positions due to different magnification and distortion, and the shown one is with the largest 
    FWHM determined by deflection fields at one of the three clumps. }
    \label{fig:betamap}
\end{figure*}

The image-plane and reconstructed source-plane UV slope maps are shown in Figure \ref{fig:betamap}. The HST image plane $\beta$ map is constructed using {PSF-matched} WFC3/F105W, F125W, F140W and F160W images which trace continua at 2125, 2513, 2805, 3091 \AA\ in the rest frame respectively. The JWST image plane $\beta$ map is constructed from {PSF-matched} NIRISS/F115W and F150W, corresponding to rest-frame continua at 2315 and 3015 \AA. NIRISS/F200W is discarded to avoid potential [\ion{O}{2}] $\lambda\lambda 3729,3726$ and [\ion{O}{3}] $\lambda 4363$ emission. The $\beta$ map constructed from JWST images exhibits more fine structures and variations, thanks to its high spatial resolution. {On both image- and source-plane $\beta$ maps, $\beta$ 
spans from -1.5 to -2.5. The bluest $\beta$ resides around the compact star clumps, reaching -2.5, caused by ionizing radiation from massive star formation. On the source plane, we confirm that $\beta$ varies on the physical scale of $<100$\,pc. }

\section{Associated Damped Ly$\alpha$ System with Super-solar Metallicity}\label{sec:dla}
% {section e.g.: DLA and HII region or just combine the 4.1 and 4.2 into 'Results' section as subsection 3.3 and 3.4 ?}

\begin{figure*}
\gridline{\fig{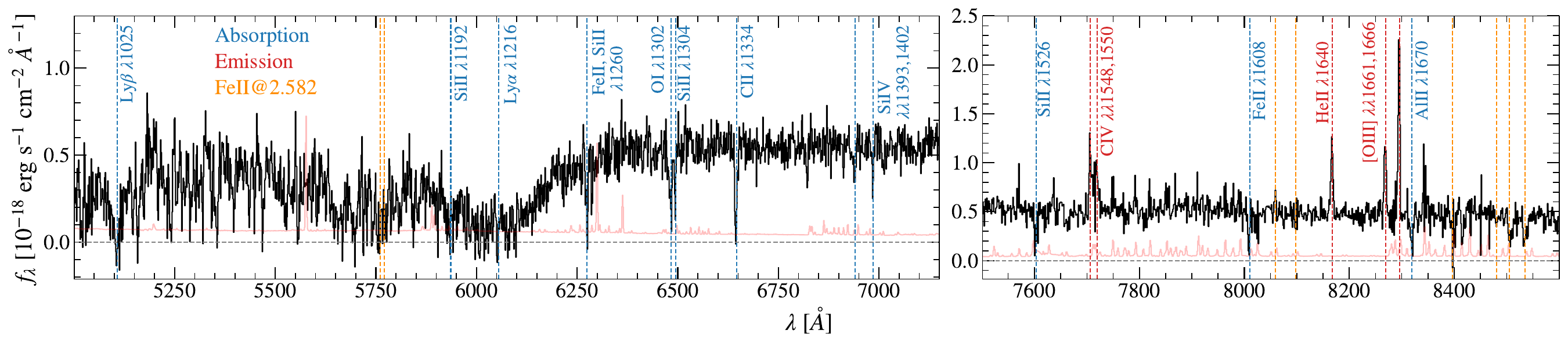}{\textwidth}{(a)}} %\label{fig:muse_spec}
\gridline{ \fig{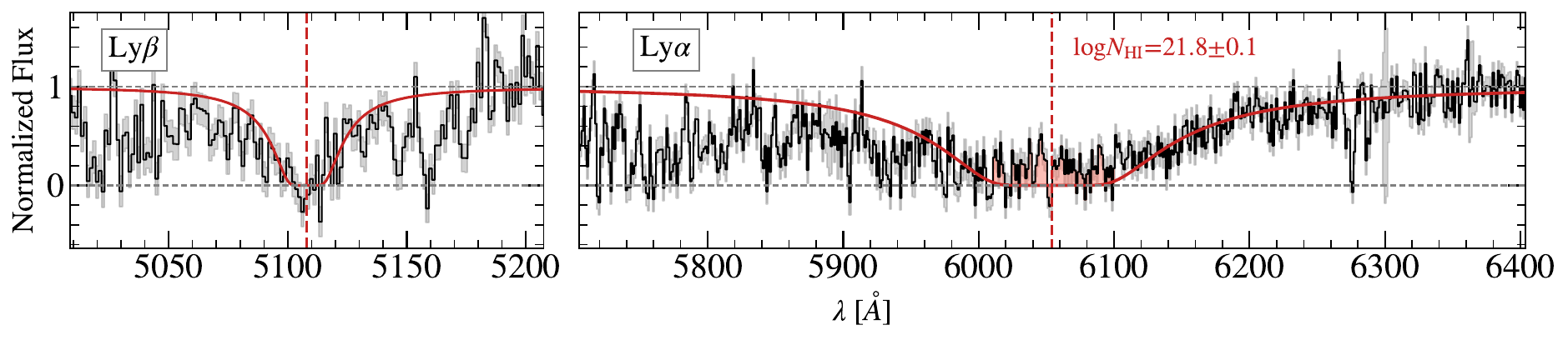}{\textwidth}{(b)}} %\label{fig:fit_dla}
\gridline{ \fig{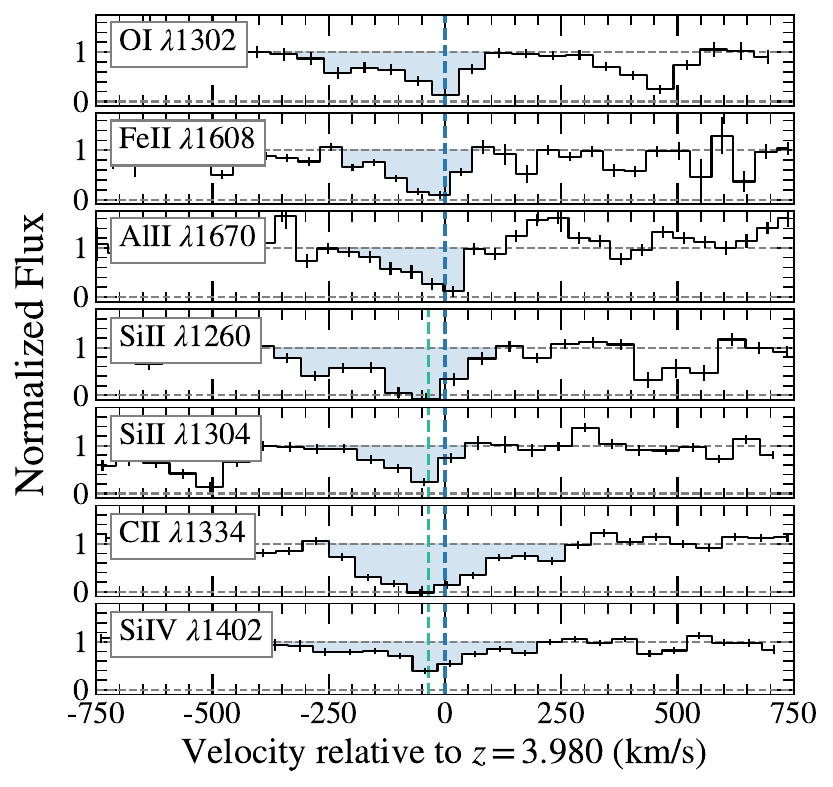}{.45\textwidth}{(c)} %\label{fig:metal}
        \fig{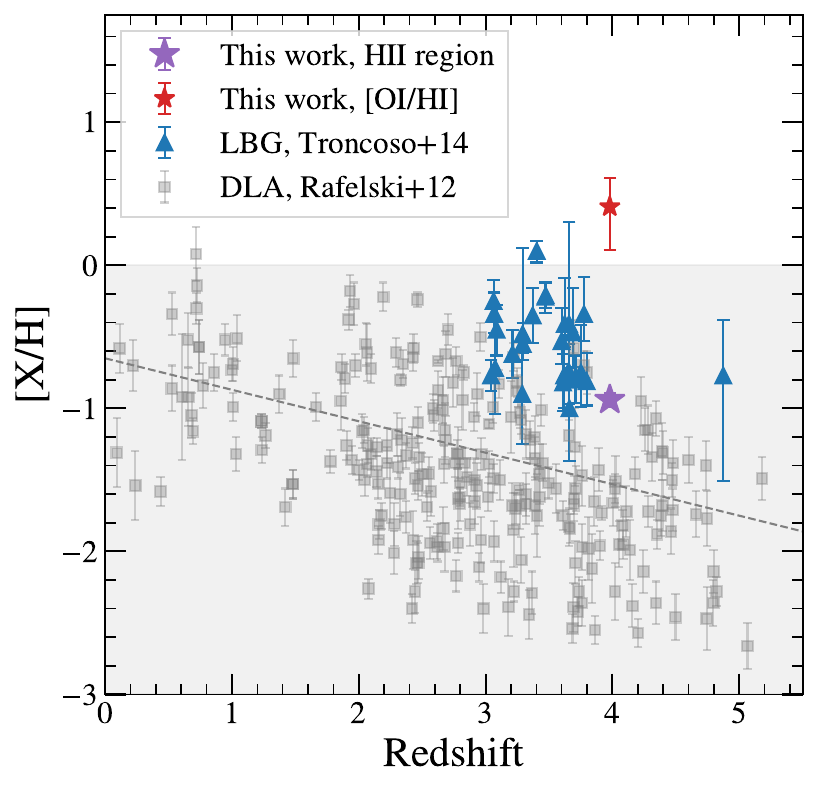}{.45\textwidth}{(d)}} %\label{fig:metal_compare}
\caption{\textbf{(a)} Rest-frame UV spectrum of A2744-arc3 extracted from 
 the VLT/MUSE 3D datacube, with the error spectrum shown in red. 
 Nebular emission lines and absorption features associated with A2744-arc3 are labelled
 with dashed red and blue lines, respectively
 A series of absorption features from an intervening \ion{Fe}{2} absorber 
 at $z=2.582$ are identified in orange. 
 \textbf{(b)} Normalized spectrum zoomed into the DLA Ly$\alpha$ and Ly$\beta$ line profiles with the Voigt profile fits, 
 with flux uncertainties indicated by gray shaded region. The red solid lines imply the best-fit result yielding 
 $\log(N_\mathrm{HI }/{\rm cm}^{-2})=21.8\pm0.1$. Tentative residual in the trough is identified by the red shaded region. \textbf{(c)} Normalized spectrum zoomed into the metal absorption line profiles. The blue shaded regions are used to measure the equivalent width (EW) of the absorbers. A tentative velocity offset of -36 km~s$^{-1}$ relative to $z=3.980$ (blue dashed) is marked in green dashed lines. \textbf{(d)} The neutral gas metallicity derived from the \ion{O}{1} absorber in compared with metal abundances of DLAs (\citealt{Rafelski2012}; gray squares) and Lyman break galaxies (LBGs) (\citealt{Troncoso2014}; blue triangles) out to $z=5$. 
 The gray dashed lines indicate the cosmic evolution of DLA metallicities in \cite{Rafelski2012}. 
 The gray shaded region marks metallicities lower than solar values. %The purple shadow ranging from $-1.39$ to $-0.94$ 
 \xj{The purple star marks the \ion{H}{2} region [O/H] abundance of A2744-arc3 derived from [\ion{Ne}{3}]/[\ion{O}{2}] empirical metallicity calibrations, as described in Section \ref{sec:emission}.}  \label{fig:Fig_metal}}
\end{figure*}

The optical spectrum of A2744-arc3, as shown in Figure \ref{fig:Fig_metal}(a), is extracted from the MUSE data cube by making a 2$\sigma$ deblended segmentation map on a 5 arcsec$\times$ 5 arcsec white-light image cutout around it. Abundant absorption and emission features are highlighted, including high-column density Ly$\alpha$ absorption line and the multiplicity of metal absorption lines (e.g., \ion{O}{1} $\lambda1302$, \ion{Si}{2} $\lambda 1304$, \ion{Si}{4} $\lambda\lambda 1393, 1402$), strong high-ionization emission lines such as \ion{C}{4} $\lambda\lambda1548,1550$, He\uppercase\expandafter{\romannumeral 2} $\lambda1640$ and \ion{O}{3}] $\lambda\lambda1661,1666$. We also detect a series of intervening \ion{Fe}{2} absorbers at $z=2.582$ along the sightline. The spectral properties of the arc are reported in Table \ref{tab}.

\begin{deluxetable*}{lcccccccccc}
% \begin{deluxetable*}{@{\extracolsep{2pt}}lcccccccccc@{}}
\tabletypesize{\small}
\tablecaption{Physical Properties of the arc derived from the photometry, absorption and emission features.  In addition, column densities and abundances of neutral gas metallicity tracers are provided.\label{tab}}
% \tablewidth{0pt}

\tablehead{\multicolumn{11}{c}{ Basic Physical Parameters}}
    \startdata
    R.A. (deg) & Dec (deg) & $z_{\rm spec}$\tablenotemark{a} & $\log M_*/M_\odot$ \tablenotemark{b} & Age (Myr) & SFR ($M_\odot$/yr) & $\log {\rm U}$ & $\mu_{\rm med}$\tablenotemark{c} & $\mu_1$\tablenotemark{c} & $\mu_2$\tablenotemark{c} & $\mu_3$\tablenotemark{c}\\
    % \hline
    3.589089 & -30.393844 & 3.980 & 7.59$^{+0.12}_{-0.10}$ & 23.95$^{+5.70}_{-8.44}$ & 1.80$^{+0.24}_{-0.21}$  & -1.34$^{+0.32}_{-0.26}$ & 45 & 142 & 46 & 30 \\
    \hline
    \hline
    \multicolumn{11}{c}{Absorption Features} \\
    \hline
      \multirow{2}{*}{Ion}& \multirow{2}{*}{$\lambda$ (\AA)}& & \multicolumn{3}{c}{CoG Method} & & \multicolumn{4}{c}{Voigt Profile Fitting} \\
     \cline{4-6}\cline{8-11}
         &   & & EW (\AA)\tablenotemark{d} & $\log$(N/cm$^{2}$)\tablenotemark{e} & [X/H]\tablenotemark{f} & & \multicolumn{2}{c}{$\log$(N/cm$^{2}$)\tablenotemark{g}} & \multicolumn{2}{c}{[X/H]\tablenotemark{h}}\\
     \hline
      \ion{H}{1}   &  1215.670 & & - & - & - & & \multicolumn{2}{c}{21.8$\pm0.1$\tablenotemark{1}} & \multicolumn{2}{c}{-}\\
      \ion{O}{1}   &  1302.168 & & 0.77$\pm$0.10 & 18.9$^{+0.2}_{-0.3}$ & \Oabund & &\multicolumn{2}{c}{18.7$\pm$0.4} & \multicolumn{2}{c}{0.21$\pm$0.40} \\
      \multirow{2}{*}{\ion{Si}{2}\tablenotemark{2}} & 1260.422 & &1.12$\pm$0.11 &- &  & & - &  \\
      & 1304.370 & & 0.47$\pm$0.10 & 17.8$_{-0.9}^{+0.4}$ &  \multirow{2}{*}{\Siabund} & & \multicolumn{2}{c}{18.0$^{+0.4}_{-0.6}$} & \multicolumn{2}{c}{\multirow{2}{*}{0.81$^{+0.36}_{-0.71}$}}\\
      \ion{Si}{4} & 1402.770 & & 0.67$\pm$ 0.11  & 17.7$\pm$0.3 & & & \multicolumn{2}{c}{17.5$^{+0.2}_{-1.6}$} \\
     \ion{C}{2}\tablenotemark{3} & 1334.532 & & 1.31$\pm$0.09 &- & -& & \multicolumn{2}{c}{-}&\multicolumn{2}{c}{-}\\
     \ion{Fe}{2}\tablenotemark{2}
     & 1608.451 & & 0.80$\pm$0.08 & 18.7$\pm 0.2$ & 1.40$\pm$0.20 & & \multicolumn{2}{c}{19.1$^{+0.2}_{-0.4}$} & \multicolumn{2}{c}{1.80$^{+0.20}_{-0.40}$} \\
     \ion{Al}{2} & 1670.787 & & 0.72$\pm$0.10 & 16.4$\pm$0.4 & 0.15$\pm$0.40 & & \multicolumn{2}{c}{16.4$^{+0.3}_{-1.3}$} & \multicolumn{2}{c}{0.15$^{+0.30}_{-1.30}$}\\
     \hline
     \hline
     \multicolumn{11}{c}{Emission Lines} \\
     \hline
     Ion & $\lambda$ (\AA) & & EW (\AA) & \multicolumn{3}{c}{FWHM (km s$^{-1}$)\tablenotemark{i}} & \multicolumn{4}{c}{Flux (10$^{-18}$ erg s$^{-1}$ cm$^{-2}$) } \\
     \hline
    \multirow{2}{*}{\ion{C}{4}} & 1548.187 & & \multirow{2}{*}{3.14$\pm$0.23} &  \multicolumn{3}{c}{118$\pm$9} &  \multicolumn{4}{c}{\multirow{2}{*}{7.8$\pm$0.6\tablenotemark{5}}}\\
     & 1550.772 & & & \multicolumn{3}{c}{$<$461\tablenotemark{4}} & \\
     \ion{He}{2} & 1640.420 & & 1.62$\pm$0.09 &  \multicolumn{3}{c}{143$\pm$8} & \multicolumn{4}{c}{3.8$\pm$0.2} \\
     \multirow{2}{*}{\ion{O}{3}]} & 1660.809 & & \multirow{2}{*}{6.25$\pm$0.27}  & \multicolumn{3}{c}{144$\pm$9} & \multicolumn{4}{c}{3.7$\pm$0.2} \\
     & 1666.150 & & &  \multicolumn{3}{c}{139$\pm$6} & \multicolumn{4}{c}{7.9$\pm$0.6}\\ 
     \hline 
     [\ion{O}{2}]\tablenotemark{6} & 3727\tablenotemark{7} & & 80.2$\pm$12.9 & \multicolumn{3}{c}{-} & \multicolumn{4}{c}{56.0$\pm$6.4}  \\
     {}[\ion{Ne}{3}] & 3868.760 & & 105.6$\pm$16.7 & \multicolumn{3}{c}{-} & \multicolumn{4}{c}{66.1$\pm$7.3}  \\
    \enddata

     \tablenotetext{a}{Redshift determined by \ion{O}{1} absorption.}
     \tablenotetext{b}{Physical parameters are derived from \texttt{BEAGLE} SED fitting.}
     \tablenotetext{c}{Magnification from the \texttt{Glafic} model. $\mu_{\rm med}$ is the median value along the arc based on the JWST/NIRISS F115W $2\sigma$ profile. $\mu_{1}$, $\mu_2$ and $\mu_3$ is the median magnification of the three mirrored clump pairs from inside (closest to the axis of symmetry/critical curve) to outside respectively.}
     \tablenotetext{d}{Rest-frame equivalent width of the absorbers. The uncertainties are estimated by varying the continuum level by 10\%, combined with the flux uncertainties.}
     \tablenotetext{e}{Column densities are derived assuming $b$=10km s$^{-1}$. Uncertainties are estimated by varying the doppler parameters $b$ from 5-15km s$^{-1}$, combined with the uncertainties from EW measurements.}
    \tablenotetext{f}{[X/H] = $\log$(X/H) - $\log$(X/H)$_\odot$. Solar (photosphere) abundances from the compilation by \cite{Asplund2009}.}
    \tablenotetext{g}{Column densities are determined by two-component Voigt profile fits to the absorption features simultaneously, assuming $b=10$km s$^{-1}$. We also perform  Voigt profile fits using  $b$=5 and 15km s$^{-1}$, and the uncertainties presented here include systematics induced by different $b$ values.}
    \tablenotetext{h}{Metallicities are calculated by summing over column densities of the two Voigt profiles.}
    \tablenotetext{i}{FWHMs are measured in the rest-frame relative to $z=3.980$, using Gaussian fits and eliminating the effect of MUSE line spread function (LSF) \citep{Bacon2017}.}
    \tablenotetext{1}{The uncertainty is determined by varying the continuum level by 10\% and the doppler parameter $b$ from 20 to 200km/s.}
    \tablenotetext{2}{The \ion{Si}{2} $\lambda$1260 and \ion{Fe}{2} $\lambda$1608 absorption might be blended with other lines.}
    \tablenotetext{3}{The \ion{C}{2} $\lambda$1334  absorption is saturated. }
    \tablenotetext{4}{The \ion{C}{4} $\lambda$1550 emission line is likely interfered by noise.}
    \tablenotetext{5}{There might be potential \ion{C}{4} absorption from neutral gas blended with its emission.}
    \tablenotetext{6}{Measured from the coadd of JWST/NIRISS slitless spectra as described in Section \ref{sec:niriss data}. We do not provide FWHMs due to blending effect in grism data and the low resolving power of NIRISS (R$\approx$150). Flux is measured by fitting single Gaussian profiles to each emission line.}
    \tablenotetext{7}{The doublet of [\ion{O}{2}]  at 3726.032 and 3728.815\AA\  cannot be distinguished.}
    
\end{deluxetable*}

\subsection{Metal-enriched \ion{H}{1} gas with high column density}

{In Figures \ref{fig:Fig_metal}(b) and \ref{fig:Fig_metal}(c), we display the DLA \ion{H}{1} and its associated metal absorption-line profiles, respectively. The metal lines exhibit tentative P-Cygni features. We use the Voigt profile to fit %all the 
absorption lines.} The Ly$\alpha$ and Ly$\beta$ absorption associated with the galaxy at $z=3.98$, depicted in Figure \ref{fig:Fig_metal}(b), indicates a \ion{H}{1} column density of $N_{\rm{HI}}=10^{21.8\pm0.1}$cm$^{-2}$, i.e.\ an extremely strong damped Ly$\alpha$ systems \citep[ESDLA, $N_{\rm{HI}}>5\times 10^{21} {\rm cm}^{-2}$;][]{not14}. \xj{The measurement is performed using \texttt{VoigtFit} \citep{VoigtFit}, \
%assuming $b=80 {\rm km\ s^{-1}}$, where the chi square converges when 
varying $b$ from 20 to 200 km s$^{-1}$. We then add the continuum uncertainty, varying the continuum level by 10\%.}  Tentative detection of residual Ly$\alpha$ emission in the absorption trough might imply partial DLA coverage. {Nevertheless, %we cannot exclude
the possibilities of contamination from %nearby or 
foreground sources or imperfect sky subtraction cannot be excluded.} 
The source-reconstruction in Figure \ref{fig:betamap} illustrates that the ESDLA extends over at least sub-kpc scales. \xj{Note that VLT/MUSE's seeing leads to a spatial resolution of ~0.6 arcsec, and the six clumps along it have radii of $\lesssim 0.2$ arcsec, beyond the spatially-resolving capability of MUSE. Thus, we did not show spatially resolved absorption 
information in this work.}

{We measure the gas-phase metallicity from the accurate indicator \ion{O}{1}. Because \ion{O}{1} and \ion{H}{1} have similar ionization potentials and are coupled by resonant charge exchange reactions, we then have [\ion{O}{1}/\ion{H}{1}] $\approx$ [O/H]. Considering the \ion{O}{1} $\lambda$ 1302 line may be saturated, we estimate the metallicity from both the Voigt profile fitting and curves of growth (CoG) analysis, \xj{by varying the Doppler parameter $b$ between 5--15 km s$^{-1}$, the typical values for the intrinsic doppler parameter of DLAs as concluded from high-resolution observations \citep[e.g.,][]{Penprase2010,Bashir2019}. We do not adopt large $b$ values (i.e. $b>$ 20 km s$^{-1}$) to avoid the underestimation of DLA metallicity without resolving the absorber subcomponent by high resolution spectroscopy \citep{pro06}.    
}%we obtain the metallicity [O/H] varying from -0.11 to 0.61. 
We list both the Voigt fitting results and the estimates from the curves of growth of metallicity in Table \ref{tab} ($b = 10$ km s$^{-1}$).} { More details about the Voigt fitting results are presented in Appendix \ref{sec:VoigtFit}.} \xj{The estimates from \ion{O}{1} and other metal lines such as \ion{Si}{2}
all yield a super-solar metallicity, with [O/H] ranging from $\approx$0.11 to 0.61 and [Si/H] from 0.21 to 1.1\ (\ion{Si}{2} + \ion{Si}{4}) respectively, corresponding to 1.29--4.07 ${\rm Z}_\odot$ and 1.63--12.59 ${\rm Z}_\odot$. ([O/H]$\approx$0.41, [Si/H]$\approx$0.74, corresponding to $Z\approx$2.57 and 5.50 ${\rm Z}_\odot$ if assuming $b=10$ km s$^{-1}$)}
% The estimates from \ion{O}{1} and other metal lines such as \ion{Si}{2}
% all yield a super-solar metallicity, with [O/H]=\Oabund\ and [Si/H]=\Siabund\ (\ion{Si}{2} + \ion{Si}{4}) respectively, corresponding to 2.57$^{+1.50}_{-1.28}$ ${\rm Z}_\odot$ and 5.50$^{+7.09}_{-3.87}$ ${\rm Z}_\odot$.
This metallicity, as highlighted in Figure \ref{fig:Fig_metal}(d), is much higher than the prediction of the cosmic metallicity evolution of DLAs ([X/H]~$\approx-1.5$ at $z=3.98$, \citealt{Rafelski2012}), and the metal abundances of typical 
star-forming galaxies at $z= 3-5$, %which are usually 
measured using nebular emission lines originated from 
their \ion{H}{2} regions  (e.g, [X/H]~$\approx -0.5$ in \citealt{Troncoso2014}). {However, we are cautious to conclude the high metallicity given the low resolution spectra (R $=$ 2000-3000) and the non-detection of weak lines such as \ion{Ni}{2} and \ion{Zn}{2}. We need high resolution spectra to confirm our measurements.}

{Additionally, we also obtain super-solar metallicity from the \ion{Fe}{2} absorption lines. Since \ion{Fe}{2} can be depleted onto the dust, the high abundance of \ion{Fe}{2} indicates not much dust in both the absorbing gas and the galaxy, consistent with the ALMA millimeter continuum non-detection. }

%We estimate the column density of low-ion absorbers \ion{O}{1} and \ion{Si}{2}, which is believed to be the best metallicity tracers for neutral gas \citep{Rafelski2012}. The results are listed in Table \ref{tab}. 
%The column density of the high-ion \ion{Si}{4} also implies a marginally consistent super-solar metallicity of $0.09^{+0.40}_{-1.20}$. Assuming two components to explain their asymmetry profiles, a double check is performed 
%by fitting two Voigt profiles to the absorption, yielding [X/H]=$0.41\pm0.20$, $0.59\pm0.40$ and $0.39\pm0.20$ for 
%\ion{O}{1}, \ion{Si}{2} and \ion{Si}{4} respectively.

\subsection{Strong narrow nebular emission -- an indication of metal-poor \ion{H}{2} region}\label{sec:emission}

{A2744-arc3 exhibits significant high-ionization nebular emission in both optical and infrared spectra, e.g., \ion{C}{4} $\lambda 1548, 1550$, \ion{He}{2} $\lambda 1640$, \ion{O}{3}] $\lambda 1661,1666$ and [\ion{Ne}{3}] $\lambda 3638$
% of \ion{C}{4} $\lambda 1548, 1550$, \ion{He}{2} $\lambda 1640$ and \ion{O}{3}] $\lambda 1661,1666$
% with narrow FWHMs and large intensity 
(see Figure \ref{fig:jwst}, \ref{fig:Fig_metal}(a) and Table \ref{tab}), which %is often 
have been detected in nearby dwarf galaxies and $z\approx1-2$ metal-poor or strong LyC emitting galaxies 
\citep[e.g.,][]{Erb2010,Schaerer2022_C4}, but not typically detected in 
%typical 
Lyman break galaxies (LBGs) %at similar redshifts 
\citep[e.g.,][]{Reddy2008}.   
These strong narrow nebular emission lines, especially \ion{He}{2} $\lambda 1640$, are believed to be tightly linked to low-metallicity galaxies with a possible contribution from Pop III-like stars \citep{Schaerer2003,Patricio2016}. % or %additional contribution from 
%X-ray binaries \citep{Schaerer2019,Fornasini2019}. 
% Nearby dwarf galaxies with minimal stellar wind features and prominent nebular emission in \ion{He}{2} and \ion{C}{4}, similar to A2744-arc3,  show low-metallicity below 12 + $\log$ O/H $<$ 8.0 ($Z/{\rm Z}_\odot< 1/5$) \citep{Senchyna}. 
% 
}

The line ratio between [\ion{O}{2}] and [\ion{Ne}{3}], two of the principal coolants in \ion{H}{2} regions, serves as a metallicity indicator for star-forming galaxies. In A2744-arc3, we obtain $\log$([\ion{Ne}{3}]/[\ion{O}{2}])$=0.07^{+0.06}_{-0.07}$, i.e.\ 12+$\log{( {\rm O/H})}=7.75\pm 0.05$, based on the empirical metallicity calibrations in \cite{Bian2018}, corresponding to [O/H] = $-0.94\pm0.05$ or ${ Z}/{\rm Z}_\odot=0.12\pm 0.01$\footnote{{%This is just a rough estimate and we discuss 
Possible systematics are discussed in Appendx \ref{sec:Ne3O2_ratio}, e.g., line blending and the scatters of the calibration relation. More observations are expected to %directly and precisely determine 
confirm its metal abundance.}}. The mass-metallicity relation (MZR) in \cite{Sanders2021} predicts 12+$\log$(O/H)=7.7$\pm$0.1 for a $M_*\approx 10^{7.6\pm0.1} {\rm M}_\odot$ galaxy 
at $z\approx3.3$; \cite{Ma2016} predicts 12+$\log {\rm (O/H)}=7.27\pm0.04$ at $z=3.98$. These values, including our SED fitting result, 
are well consistent with models of \ion{He}{2} emitters, which favour
sub-solar metallicities ($\sim0.1~{\rm Z}_\odot$) and young stellar ages ($10^{7}-10^{8}$ years; \citealt{Saxena2020}). \xj{A further evidence is \ion{C}{4} $\lambda\lambda$ 1548,1550 in pure emission, which is revealed uniquely in galaxies with 12+log(O/H)$\lesssim$8 \citep{Mingozzi2022}}
The line ratio of \ion{O}{3}] $\lambda$1666/\ion{He}{2} $\lambda$1640  is $2.08\pm0.19$, 
inconsistent with any photoionization models powered by AGN or shocks which expect more flux in
\ion{He}{2} $\lambda$1640 than in \ion{O}{3}] $\lambda$1666 ($\log$(\ion{O}{3}] $\lambda$1666/\ion{He}{2} $\lambda$1640) $\approx -3$ to $-0.5$) \citep{Allen2008,Feltre2016,Mainali2017,Senchyna2017}.
This disjoint favours the explanation that the
nebular emission in A2744-arc3 should primarily be powered by photoionization due to massive stars \citep{Gutkin2016}. 
%({\bf cite a reference indicating the line ratio consistent with stars}). %rather than AGN.

\bigskip

\section{Discussion}\label{sec:discussion}

\subsection{A2744-arc3 has properties similar to those local analogues}\label{sec:analogue}

SL2S J0217 \citep{James2014} and SBS 0335-052 \citep{Brammer2012,Berg2018,Erb2019} at lower redshifts share %very 
similar features to A2744-arc3. Strong nebular emission lines such as \ion{C}{4} $\lambda\lambda 1548,1550$, \ion{He}{2} $\lambda 1640$ and \ion{O}{3}] $\lambda\lambda 1661,1666$ are notable in their UV spectrum, suggesting a very high ionizing field from the low-metallicity starbursting region and young stellar populations. 
SBS 0335-052 is a nearby blue compact dwarf galaxy ($z=0.0009$) which shows a damped Ly$\alpha$ profile of $\log{N_{\rm{HI}}/{\rm cm^{-2}}}=21.70\pm0.05$, while SL2S J0217, a low-mass ($M<10^9\ {\rm M}_\odot$), low-metallicity (${ Z}<1/20\ {\rm Z}_\odot$) galaxy at $z=1.8$ 
magnified by a foreground massive galaxy, may be characterized by a superposition of strong Ly$\alpha$ emission and 
damped absorption. SL2S J0217 also shows a blue UV slope of $\beta=-1.7\pm0.2$ over rest-frame 2100-2800 \AA, 
and an ionization parameter of $\log{ U}\approx -1.5$.

In addition, Green Pea galaxies at $z<1$  %compact luminous 
%galaxies 
with extremely high degree of ionization
in \ion{H}{2} regions often present a blend of strong emission and
underlying absorption \citep{McKinney2019}, with column densities 
ranging from $\log{N_{\rm HI}/{\rm cm^{-2}}}=19-21$, well above the LyC thick limit \citep[e.g.,][]{Steidel2018}. The presence of the DLA in the spectra of A2744-arc3 at $z=4$ 
further 
%This 
suggests that low-mass, low-metallicity galaxies, even at high redshift of $z\gtrsim4$, %with hard ionization and metal-poor population of stars  
may commonly possess abundant neutral hydrogen gas,  %with a wide range 
%of column densities, 
and the ultimate Ly$\alpha$ and ionizing photon escape may strongly depend on geometrical effects.

\bigskip

\subsection{The metal-enriched neutral gas reservoir around young, low-mass and metal-poor Galaxies}

{{As discussed above, A2744-arc3 is a low-mass ($M_*\approx10^{7.6}\ {\rm M}_\odot$) galaxy with low-metallicity (${ Z} \approx 0.1\ {\rm Z}_\odot$) \ion{H}{2} regions},  % in terms of the metallicity
estimated from the nebular emission lines. In contrast, metal absorbers associated with the DLA 
%yields a {possible} 
shows that the HI gas around the galaxy may have the super-solar metallicity.} %for the \ion{H}{1} gas.}
Super-solar metallicity of the neutral gas around a low-Z, high-U and low-mass galaxy is also 
reported in \cite{Berg2018}. The metal-enriched gas can be generated by a high efficiency in 
the gas cooling and enriched by the recycled inflow from the star formation \citep[e.g.,][]{Frye2019}. %This suggests that although {\bf stellar populations in the} low-mass galaxies could be 
%young, with age of $\approx 10^7$ years. % (? double check), 
Due to the metal-enrichment process is rapid, the metal-enriched environment may be
further enhanced by the continuous star formation in the high-redshift Universe. %This implies that Pop-III star dominated galaxies 
%may be found in even younger or lower mass galaxies \citep{magg2018}. 
%{\bf would be best to give a reference}f

\begin{figure*}[htbp]
    \centering
    \includegraphics[width=\textwidth]{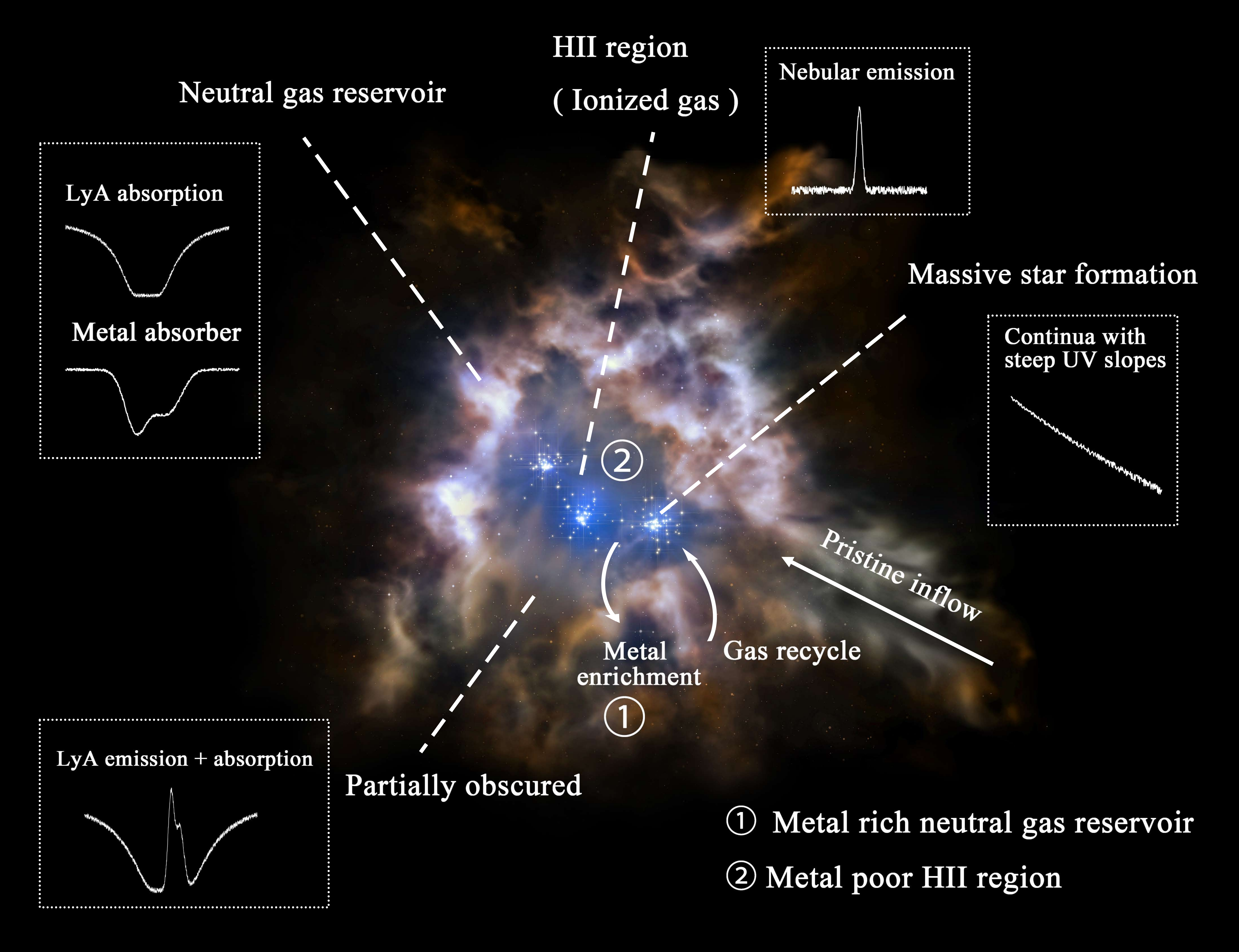}
    \caption{The schematic toy model of A2744-arc3 as a low-mass, low-metallicty galaxy with high metallicity neutral gas reservoir. The blue stellar clumps inside mark the compact and bursty star-forming regions in A2744-arc3, on scales of sub-kpc.  Their hard ionization field forms hot, thin and highly-ionized \ion{H}{2} regions around them, where strong narrow nebular emission lines are produced. The gas outer layer denotes the neutral gas reservoir with varying column densities in the CGM. Its geometry determines the shape of Ly$\alpha$ emission/absorption we observed. When sightlines are blocked by high-column density gas, damped Ly$\alpha$ absorption is observed; in partially obscured case, Ly$\alpha$ emission with underlying absorption is observed. Three arrows mark possible gas cycles within the galaxy. Feedback from central star-formation, such as stellar winds, could metal-enrich the neutral gas, leading to strong metal absorbers with P-Cygni profiles. The recycled neutral gas %, as well as potential pristine inflow from cosmic filaments, 
    can further efficiently fuel the next-generation star formation.
    %{\bf add more descriptions} 
    }
    \label{fig:diagram}
\end{figure*}

%Including the pristine inflow from the IGM \citep[e.g.,][]{Daddi2021}, 
We propose a simple schematic model to explain the {metallicity variances in the ISM/CGM and star-forming region at $z\approx4$}, as shown in Figure \ref{fig:diagram}. Massive stars produce a hard radiation field, forming a highly ionized \ion{H}{2} region. %The environment can be 
%metal enriched %during intense forming activities, for example by 
%due to stellar winds from young stellar populations {\bf References}. 
As mentioned above, the compact star-forming regions could be surrounded by  \ion{H}{1} clouds with varying column densities \citep{Erb2019}. %({\bf references}). 
%Damped Ly$\alpha$ absorption 
DLA can be observed when the sightline is completely blocked by high column density neutral clouds, %with high column densities, 
while %there would remain 
residual Ly$\alpha$ emission in the absorption trough can be detected in partially obscured cases. These outskirt environments
could be metal-polluted by the efficient feedback from massive stars %\ion{H}{2} region, 
due to the high-velocity stellar wind. %from massive young stellar populations. 
The metal-enriched gaseous environment could result in the presence of 
metal absorbers detected in Figure \ref{fig:Fig_metal}(c), 
sometimes with the detectable P-Cygni profiles \citep{Erb2012}. This scenario explains the tentative detection of Ly$\alpha$ emission shown in Figure \ref{fig:Fig_metal}(b) and asymmetry profiles of metal %super-solar metallicity 
absorbers shown in Figure \ref{fig:Fig_metal}(c). 

\xj{Figure \ref{fig:Fig_metal}(d) implies that the metal abundance of A2744-arc3 is tentatively higher than that predicted by 
cosmic evolution of DLA metallicities. This could be 
resulted from strong feedback and metal-enrichment. Whether the enhanced metal enrichment is prevalent among low mass galaxies, more observations on DLA host galaxies are required. }

\subsection{Implications of the spatially resolved UV slope map}
{The ultraviolet (UV) continuum slope $\beta$ ($f_\lambda \propto \lambda^{\beta}$) is a prominent metric for properties of high-redshift galaxies. Both simulations and observations suggest that,  
bluer galaxies observed at $z = 3-4$ have $\beta = -2$, 
while values as blue as $\beta = -3$ can in principle be produced by 
a young (age $\lesssim10^8$ years), low-metallicity (${ Z}\lesssim10^{-3}{\rm Z}_\odot$) stellar population \citep[e.g.,][]{Schaerer2003,Jiang2020,Topping2022}. \citet{Bouwens2010} predicted that the bluest $\beta$ for low metallicity ($0.02 - 0.2\ {\rm Z}_\odot$) starbursts is about $-2.7$, 
consistent with $\beta$ around the three clumps in Figure \ref{fig:betamap}. This further indicates the low-metallicity nature of these early star clusters. 
Aside from the blue radiation from young stars, recombination continuum radiation from the ionized gas can make the UV spectrum redder. When metallicity is greater than 0.02~${\rm Z}_\odot$, the UV slope generated by the recombination continuum become greater (redder) than $-2.5$.  With a high spatial resolution, we are  able to separate the relatively red nebular continuum emission from the ionized gas from the massive star continuum \citep[e.g.,][]{Bouwens2009} which may be consistent with what we have observed 
here in A2744-arc3 with JWST. 
% When metallicity is greater than $0.02{\rm Z}_\odot$, then the UV slope generated from the nebular light part become greater than the value of -2.5
}

\subsection{Implication of such a Low-mass, but dominating population of galaxies during  Reionization}

%{\bf will edit later}

%A2744-arc3 at $z=4$ exhibits a hard radiation with $\beta$ reaching $-2.5$ from young metal-poor stellar populations. 
%The stellar mass of A2744-arc3 is about $10^{7.6} {\rm M}_\odot$, with the \ion{H}{2} region metallicity of $\approx0.12\ {\rm Z}_\odot$. 

The UV luminosity of A2744-arc3 is $M_{\rm UV}=-17.8\pm 0.1$, corresponding to 0.049 $L_{\rm{UV}}^*$ galaxies 
in the faint-end of the galaxy luminosity function \citep[$M^*=-20.88\pm0.08$,][]{Bouwens2015}. 
% The strong lensing plus JWST excellent spatial resolution 
% allow us to resolve this galaxy on $\approx100$ pc, making it possible to study the spatial variation of properties, 
% including the UV radiation, ionization parameter and metallciity for a $z=4$ galaxy. 
Through the spatially-resolved emission and absorption lines, 
A2744-arc3 at $z=4$ provides us with a text-book example, allowing 
an unparalleled view of low-mass galaxies at the epoch of Reionizaiton (EoR).  

%which makes it an ideal template for studies on reionization-era sources. 
Low-mass, metal-poor galaxies with hard ionizing radiation show much promise to dominate the ionizing photon budget during EoR \citep{Wise2014}, with a predicted escape fraction of $\sim 20\%$ \citep{Duffy2014}. 
Similar to lower-redshift analogs \citep[e.g.,][]{James2014,Berg2018,Erb2019}, A2744-arc3 at $z=4$  
has shown the prevalence of a neutral gas reservoir. %in such galaxies. 
Given these observations, 
we argue that, when modelling the escape of ionizing radiation into the intergalactic medium during EoR, the neutral gas reservoir surrounding galaxies themselves (see Figure \ref{fig:diagram}) should be seriously considered.  
Our identification of the metal-enriched, neutral-gaseous clouds might support the picture that only a small fraction of ionizing radiation of low-mass galaxies escape through the circumgalactic channels with low column density of neutral gas. 
With more observations of such galaxies, the statistics of the spatial variations of ionization, metallicity, HI covering fraction can be 
better constrained. 
In particular, the distribution of the optically thick HI clouds could 
affect the photon escape by geometrical effects. Then, these HI clouds could 
regulate the star formation and ionization fields through feedback. 
On the one hand, the outflow from stellar winds could metal enrich the clouds, and then, the 
clouds could further power the star formation through the recycled inflow \citep[e.g,][]{Frye2019}.
%such as metal-enrich outflow and neutral gas inflow 
% ({\bf also see xxxx}). 
In the future, with telescopes such as the Large Ultraviolet Optical Infrared Surveyor (LUVOIR; \citealt{Roberge2018}), we will be able to better constrain the spatial distribution of optically-thick HI absorbers at 
$z=2-5$, and improve our understanding towards the UV photon escape of low-mass galaxies in the high-redshift Universe.

\section{Conclusion}
In this letter, we presented a detailed analysis on a strongly magnified arclet, A2744-arc3, in the Abell 2744 field with the combination of  JWST/NIRISS (both imaging and grism spectroscopy), HST/WFC3, HST/ACS, VLT/MUSE and ALMA observations. The photometry and SED fitting imply that A2744-arc3 has a young age of $ 24^{+6}_{-8}$ Myr, high-ionization of $\log{ U}$ of $-1.3\pm0.3$, low-mass of $\approx 10^{7.6\pm0.1} {\rm M}_\odot$, and the low-metallicity \ion{H}{2} regions of 12+$\log{\rm (O/H)}\approx7.75$, corresponding to 0.12 ${\rm Z}_\odot$ at $z=3.98$.

The reconstructed source-plane images based on JWST observations provide %unprecedented 
the resolved UV slope at the spatial resolution of $\approx$100 pc. 
Three compact clumps show blue UV slopes, ranging approximately from $-1.5$ to $-2.5$, consistent with  \cite{Vanzella2022}, implying %claimed
%that the three are 
young massive star clusters with bursty star formation activities. 

%Abundant emission and absorption features in the extracted MUSE spectrum 
%Through both emission and absorption studies, 
%allow us to study the 
%we analysed the properties of ionized and neutral gas in its interstellar and circumgalatic medium. 
Through absorption studies, we have identified an associated extremely strong DLA system, with $\log N_{\rm HI} = 21.8\pm0.1$. % with tentative detection of residual Ly$\alpha$ emission in its dark trough, indicative of a partial DLA coverage. 
Multiple metal absorbers indicate that this DLA may be super-solar metallicity, \xj{
with [O/H] varying from 0.11 to 0.61 assuming the doppler paramter $b$ in the range of 5--15 ${\rm km~s^{-1}}$,}
%\Oabund assuming the doppler paramter $b\approx$10 ${\rm km~s^{-1}}$,
higher than typical DLAs and LBGs at similar redshifts, indicative of strong metal-enrichment from low-mass galaxies. Additionally, the tentative P-Cygni profiles of the metal lines strengthen evidences for the young stellar feedback. {To confirm our conclusion, we need higher resolution spectra with further larger wavelength coverage to measure the absorbing gas-phase metallicity.} A2744-arc3 also shows strong narrow nebular emission lines of \ion{C}{4} $\lambda\lambda 1548, 1550$, \ion{He}{2} $\lambda 1640$ and \ion{O}{3}] $\lambda\lambda 1661,1666$, indicative of a metal-poor \ion{H}{2} region with a hard ionization field.

Combing the lower-redshift analogs SL2S J0217 at $z=1.8$ and SBS 0335-052 at $z=0.0009$, our observations of A2744-arc3 at $z=4$ suggest that neutral gas reservoir may be prevalent around low-mass galaxies. 
%for extremely ionizing and metal-poor populations. 
%To explain it, we propose a toy model  
In our schematic model (Figure \ref{fig:diagram}), 
massive star formation reside within a metal poor \ion{H}{2} region. The galaxy is surrounded by a large reservoir of neutral gas with a metallicity that could reach solar metallicity.  %metal-enriches outskirt neutral gas through feedback. 
The %escape of 
ionizing photons %is determined by the geometry of gas clouds of varying column densities. 
could only escape through low column density channels. 
A2744-arc3 at $z=4$ provide us an excellent picture in understanding % as an excellent reference to 
 faint-end galaxies that may have played a vital role during the EoR. 
%We can just stop here. 

%Further observations are expected to probe the interplay between the metal-poor ionizing gas and metal-rich neutral gas. JWST/NIRSpec IFU would provide us a brandnew insight into the gas kinematics of \ion{H}{2} region. XXXX, with extreme high spatial resolution, would allow us to explore the spatial variation of the neutral gas, \ion{H}{1} covering fraction and Ly$\alpha$ escape fraction.  XXXX

\section*{Acknowledgments}

Xiaojing Lin and Zheng Cai thank Eros Vanzella, Lilan Yang, Dandan Xu and Shiwu Zhang for very helpful discussions.  X.L. thanks Zechang Sun for the useful tool \texttt{SpecViewer} \citep{ZechangSun2021}.
Zheng Cai, X.L., Y.W., Z.L., M.L.,\& S.Z. are supported by the National Key R\&D Program of China (grant no.\ 2018YFA0404503), the National Science Foundation of China (grant no.\ 12073014), and the science research grants from the China Manned Space Project with No. CMS-CSST2021-A05.
F.S.\ acknowledges support from the NRAO Student Observing Support (SOS) award SOSPA7-022. 
F.S.\ and E.E.\ acknowledge funding from JWST/NIRCam contract to the University of Arizona, NAS5-02105. J.X.P. acknowledges partial support from NSF AST-2107991. D.E. acknowledges support from the Beatriz Galindo senior fellowship (BG20/00224) from the Spanish Ministry of Science and Innovation, projects PID2020-114414GB-100 and PID2020-113689GB-I00 financed by MCIN/AEI/10.13039/501100011033, project P20\_00334  financed by the Junta de Andaluc\'{i}a, and project A-FQM-510-UGR20 of the FEDER/Junta de Andaluc\'{i}a-Consejer\'{i}a de Transformaci\'{o}n Econ\'{o}mica, Industria, Conocimiento y Universidades.

This work is based on observations made with the VLT/MUSE, ALMA, NASA/ESA Hubble Space Telescope and  NASA/ESA/CSA James Webb Space Telescope. MUSE observations collected at the European Organisation for Astronomical Research in the Southern Hemisphere under ESO program 094.A-0115. HST and JWST data were obtained from the Mikulski Archive for Space Telescopes at the Space Telescope Science Institute, which is operated by the Association of Universities for Research in Astronomy, Inc., under NASA contract NAS 5-03127 for JWST and NAS 5–26555 for HST. The JWST observations are associated with program ERS-1324. The HST observations are associated with the Hubble Space Telescope Frontier Fields program. The authors acknowledge the GLASS team for developing their observing program with a zero-exclusive-access period. The following ALMA data is also adopted: ADS/JAO.ALMA\#2018.1.00035.L and \#2013.1.00999.S. ALMA is a partnership of ESO (representing its member states), NSF (USA) and NINS (Japan), together with NRC (Canada), MOST and ASIAA (Taiwan), and KASI (Republic of Korea), in cooperation with the Republic of Chile. The Joint ALMA Observatory is operated by ESO, AUI/NRAO and NAOJ. The National Radio Astronomy Observatory is a facility of the National Science Foundation operated under cooperative agreement by Associated Universities, Inc.

\vspace{5mm}
Some of the data presented in this paper were obtained from the Mikulski Archive for Space
Telescopes (MAST) at the Space Telescope Science Institute. The specific observations analyzed
can be accessed via \dataset[10.17909/T9KK5N]{https://doi.org/10.17909/T9KK5N} and
\dataset[10.17909/91zv-yg35]{https://doi.org/10.17909/91zv-yg35}.

\facilities{JWST/NIRISS, ALMA, VLT/MUSE}
\software{JWST calibration pipeline \citep[1.6.2][]{Bushouse2022}
    BEAGLE \citep{Chevallard2106},
        VoigtFit \citep{VoigtFit},
        SpecViewer\citep{Sun2022}}

\appendix

\section{Systematics of lensing models}\label{sec:lensing model systematics} 
\subsection{Systematics on photometry}
\begin{figure}
    \centering
    \includegraphics[width=0.5\textwidth]{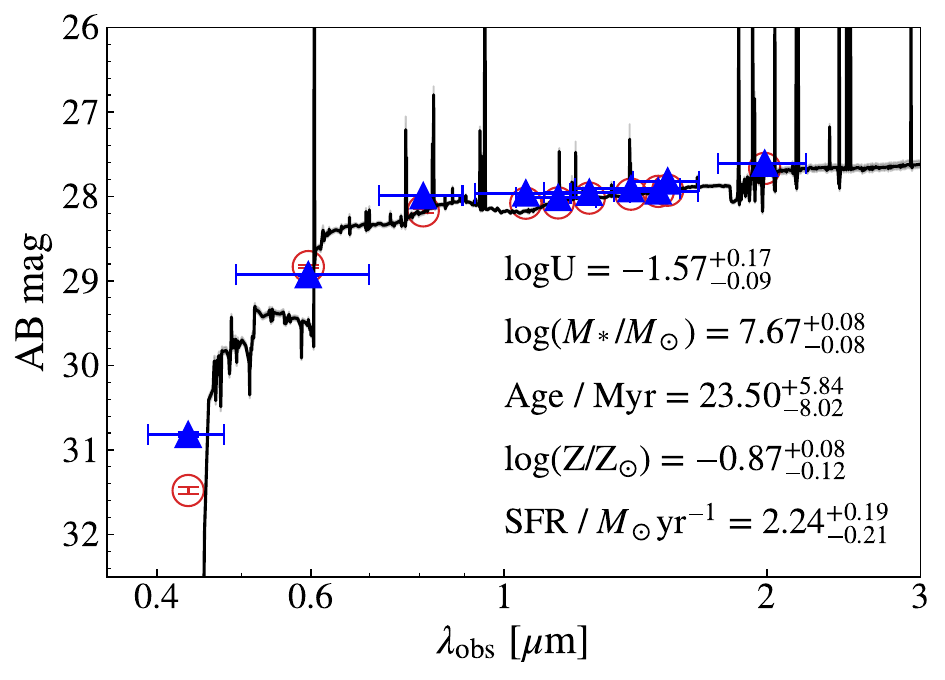}
    \caption{Photometry and SED analysis of A2744-arc3, with delensed fluxes measured using \citet{Richard2021} lensing model. The blue triangles denote the measurement and the red circles the best-fit result.  Although the adopted lensing model is different, this \citetalias{Richard2021} results are well consistent with that shown in Figure \ref{fig:jwst}.}
    \label{fig:SED_R21}
\end{figure}

In Section \ref{sec:photometry} we measure the delensed-flux of A2744-arc3 using \texttt{Glafic} lensing model. Here we repeat the measurement using the lensing model in \citet[hereafter \citetalias{Richard2021}]{Richard2021}, and the physical properties of A2744-arc3 keeps consistent with those shown in Figure \ref{fig:SED_R21}, with the difference of the stellar mass $\approx 1\%$ and age $\approx 2\%$.

\subsection{Systematics on source reconstruction}
In Section \ref{sec:reconstruction} we adopt \texttt{Glafic} lensing model to reconstruct the intrinsic UV slope distribution of A2744-arc3. Limited by the lensing model, we can hardly reconstruct the three clump pairs  onto the source plane simultaneously, where each pair should have perfectly overlapped. Among all publicly released lensing models, the \texttt{Glafic} model presents the smallest offsets between source-plane positions of these clumps pairs, yielding 0\farcs010, 0\farcs017, and 0\farcs033 from inside to outside respectively. To avoid model-dependent systematics introduced by the non-overlap, the reconstruction is only performed on the west side of the arclet that is predicted to have larger separation for the three clumps.

\begin{figure*}[htbp]
    \centering
    \includegraphics[width=\textwidth]{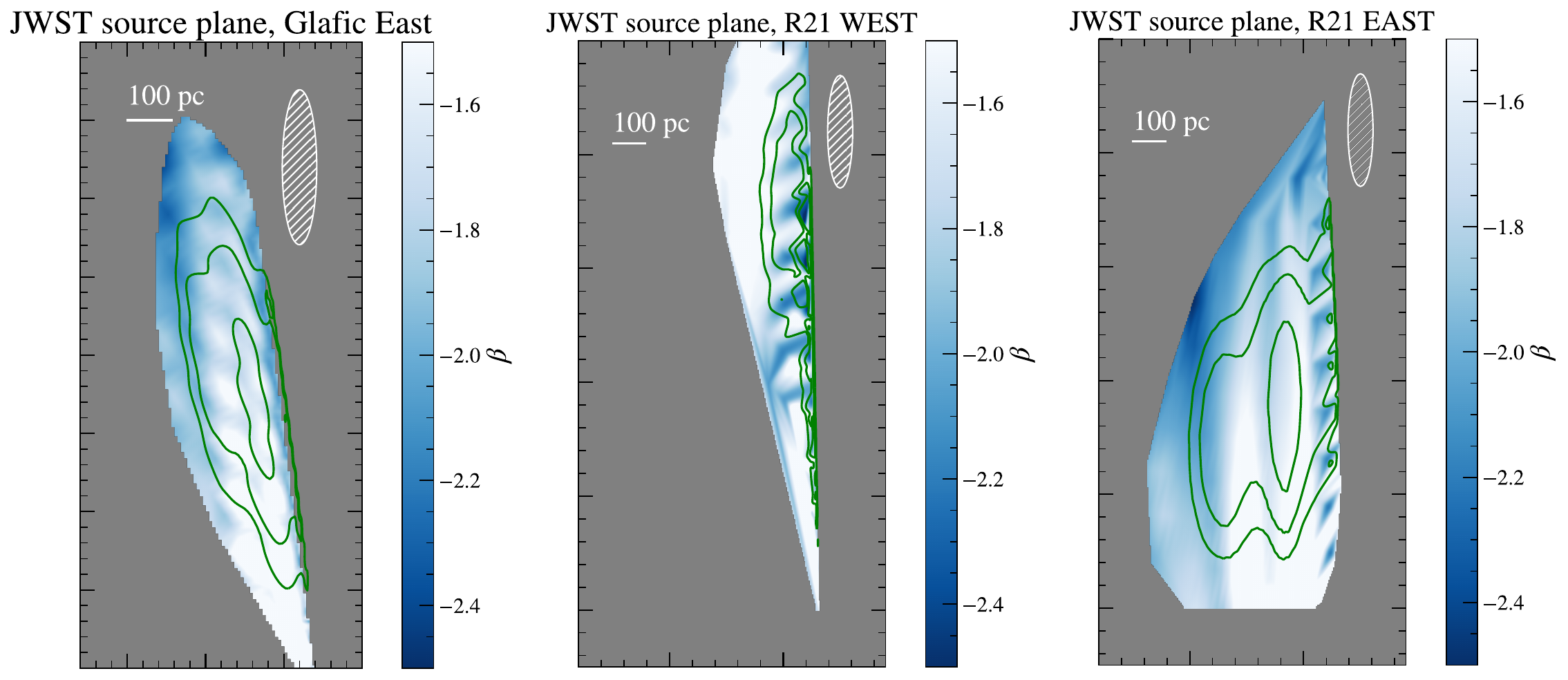}
    \caption{Reconstructed UV slope distribution on the source plane using JWST/NIRISS images. The left panel shows the reconstruction for the east side of A2744-arc3 using the \texttt{Glafic} model. The middle presents the reconstruction for the west side of A2744-arc3 with the \citetalias{Richard2021} model and the right for its east part. See more details in Figure \ref{fig:betamap}.}
    \label{fig:lensing model reconstruction}
\end{figure*}

Different lensing model yields different magnification and deflection fields, which would lead to different reconstructed morphology. As discussed before, lensing models can not perfectly recover the symmetric clumps in A2744-arc3 in the source plane, and we choose \texttt{Glafic} because it provides smallest offsets for the reconstructed clump pairs. In Figure \ref{fig:lensing model reconstruction} we show the reconstructed UV slope using the \citetalias{Richard2021} lensing model, which yields offsets of  0\farcs12, 0\farcs05 and 0\farcs19 for the three clump pairs, as well as the \texttt{Glafic} reconstruction for the arc on the east side.  Though the source-plane morphology and layout of the three clumps varies with different lensing models, the UV slope keeps within in the range $-2.5$ to $-1.5$ and the spatial variation keeps on scales of $<$100 pc.

\section{Sysematics on metallicity}\label{sec:Ne3O2_ratio}
{ In Section \ref{sec:emission} we estimate the \ion{H}{2} region metallicty of A2744-arc3 using the [\ion{Ne}{3}]/[\ion{O}{2}]-metallicity callibration relation in \citet{Bian2018}:
$12+\log (\mathrm{O} / \mathrm{H})=7.80-0.63 \times \log ([\mathrm{NeIII}] /[\mathrm{OII}])$. Due to the low resolving power of JWST/NIRISS ($R\approx 150$), [\ion{Ne}{3}] at 3869\AA\ is blended with \ion{He}{1}+ H8 at 3889\AA\ \citep[e.g.,][]{Schaerer2022}, so that the true [\ion{Ne}{3}]/[\ion{O}{2}] ratio would be lower and hence the metallicty a bit higher than the [\ion{Ne}{3}]/[\ion{O}{2}]-derived one. But this won't change the metal-poor nature of A2744-arc3. 
Despite large scatters in the [\ion{Ne}{3}]/[\ion{O}{2}]-metalicity relation \citep[e.g.,][]{Schaerer2022} concluded from various galaxies at $z=0-8$, recent observations show that the metalicity of high-redshift galaxies \citep{Sanders2020,Curti2022} tends to be consistent with or smaller than the prediction from the empirical calibration in \citet{Bian2018}. The latter puts A2744-arc3 to the lower metallicty end. All these facts, along with the strong UV emission lines, support that A2744-arc3 should possess a metal-poor \ion{H}{2} region.  A more precise measurement of its metallicity is expected in the future with high resolution spectroscopy. }

\section{Two-component Voigt profile fit}\label{sec:VoigtFit}

\begin{figure*}
    \centering
\gridline{
\fig{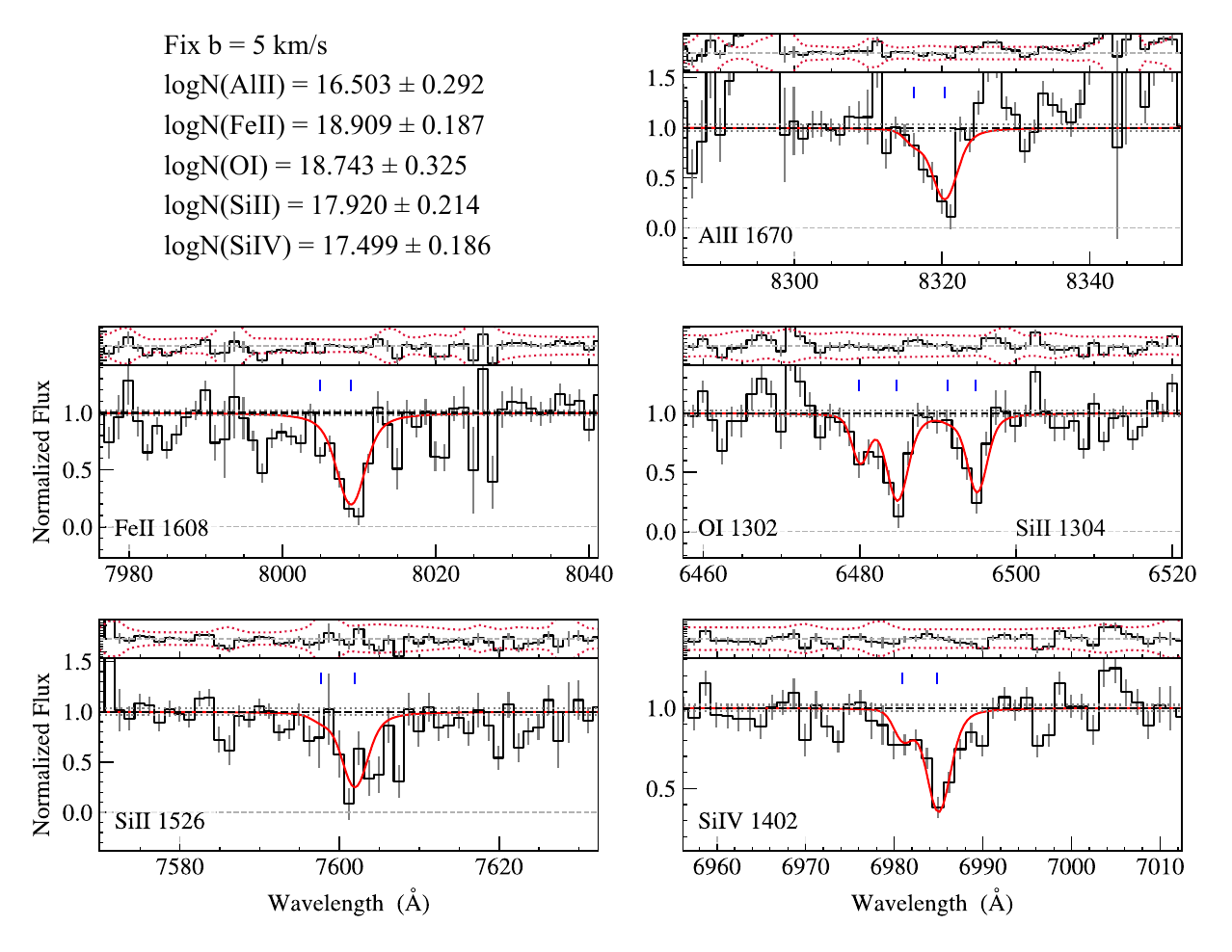}{0.49\textwidth}{(a)}
\fig{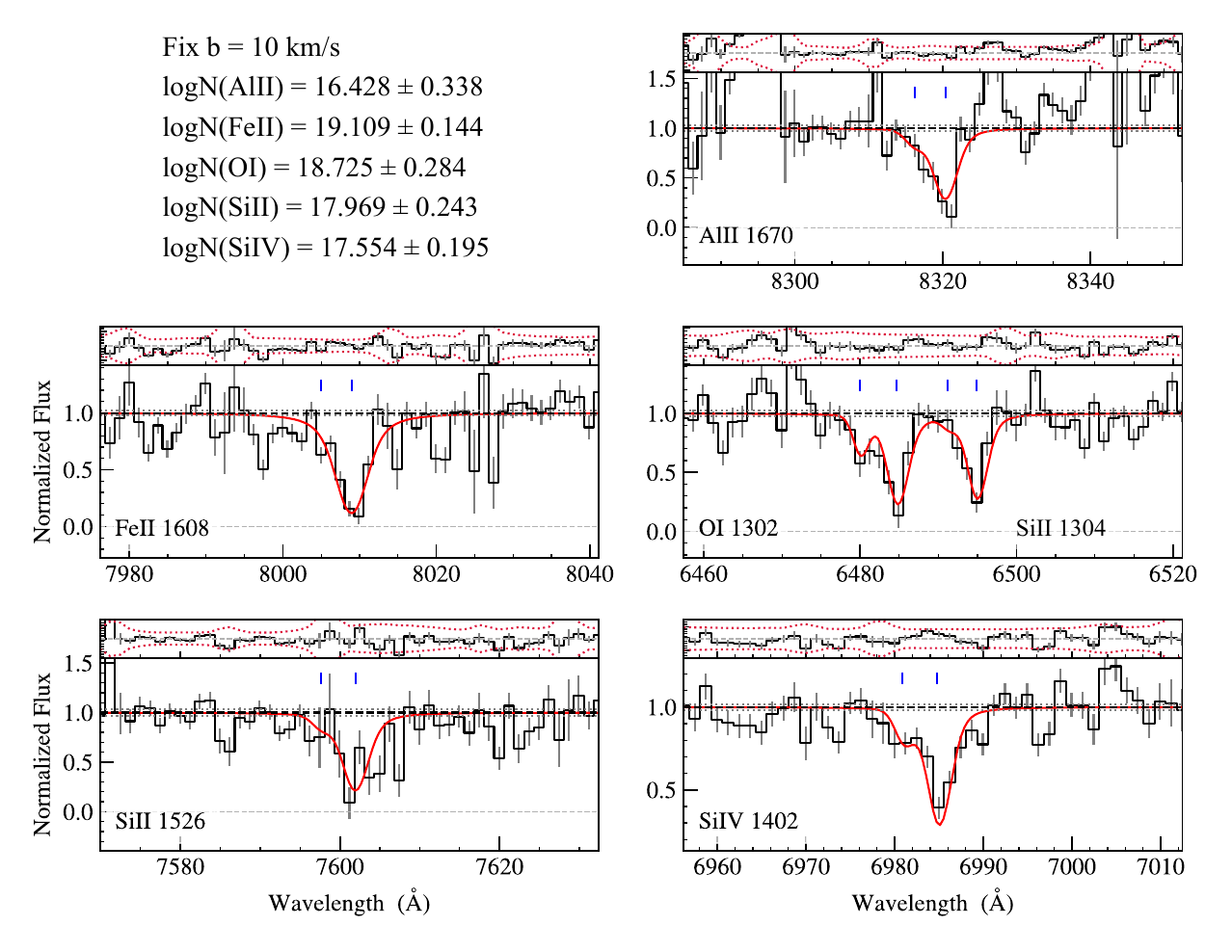}{0.49\textwidth}{(b)}
}
\gridline{
\fig{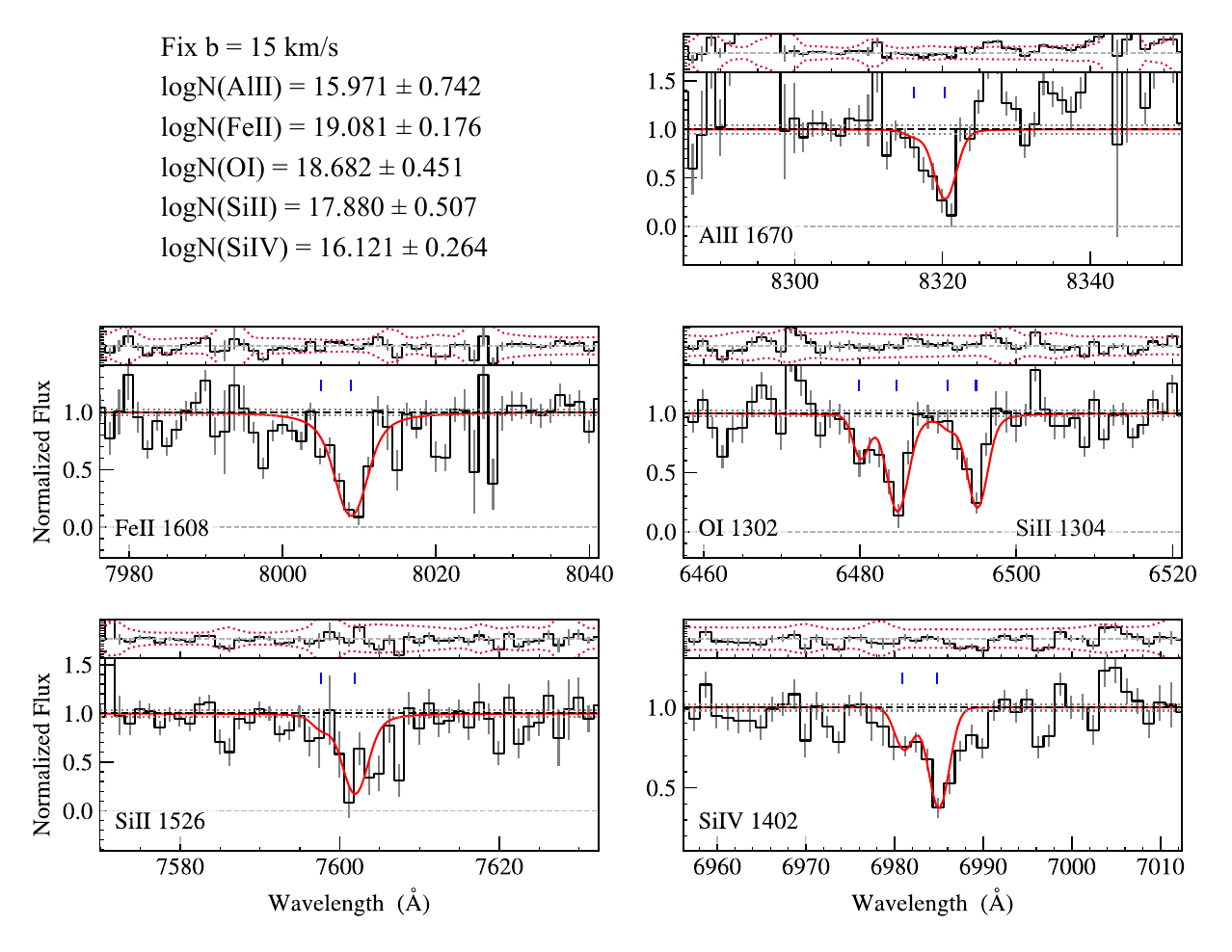}{0.49\textwidth}{(c)}
\fig{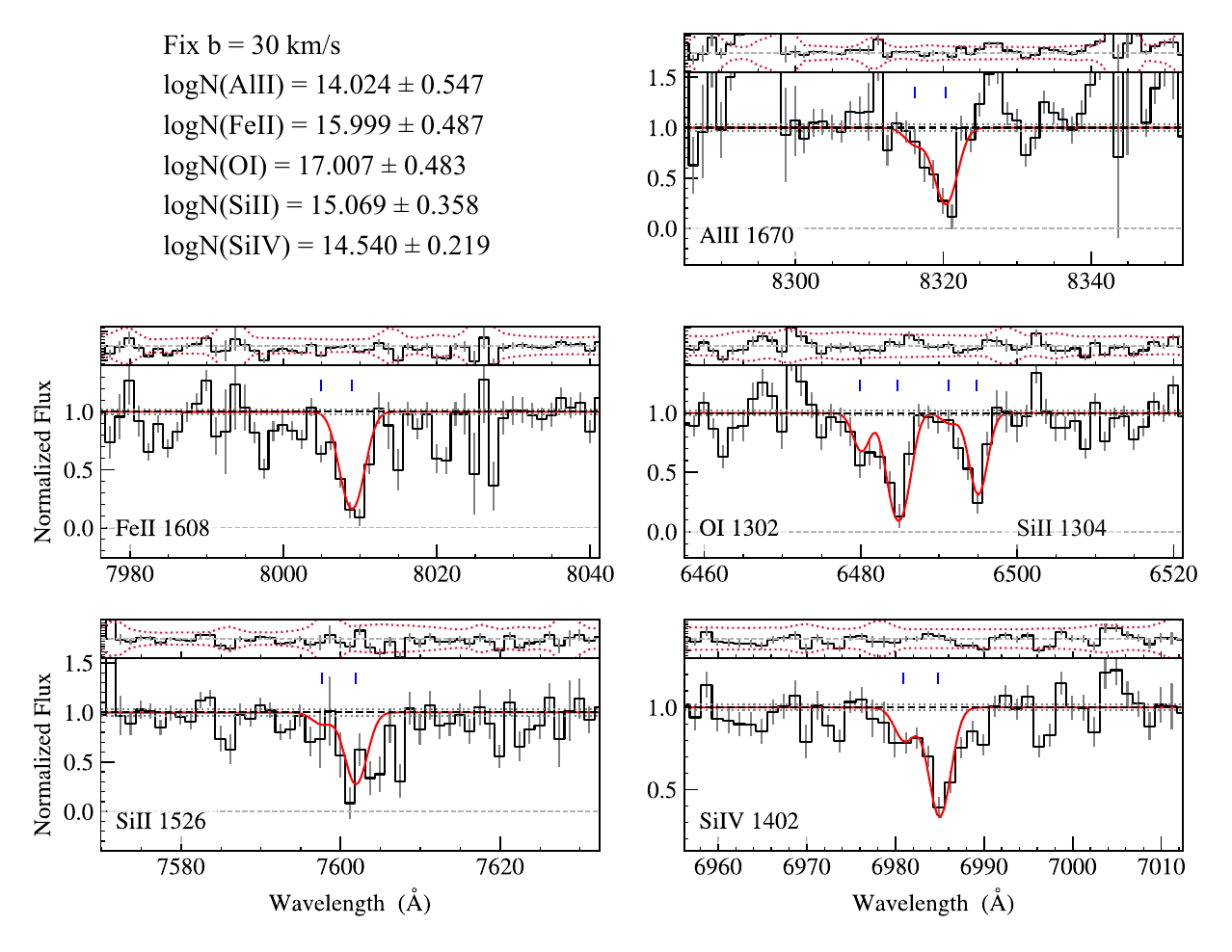}{0.49\textwidth}{(d)}
}
\caption{Voigt profile fits to the \ion{Al}{2} $\lambda$1670, \ion{Fe}{2} $\lambda$1608, \ion{O}{1} $\lambda$1302, \ion{Si}{2} $\lambda$1526 and \ion{Si}{4}  $\lambda$1402 absorbers, assuming two components for each absorbers and fixing the doppler parameters $b$ to 5, 10, 15 and 30 km s$^{-1}$. Red lines are the best-fit profiles and the blue vertical lines mark the positions of the two components for each absorber. The top row of each panel shows residuals of the best-fit profiles (black) and 3-$\sigma$ uncertainties (red dashed).  The total column density of each absorber, obtained by summing over their two components, is labelled, in units of logN/cm$^{-2}$.  \label{fig:fix_b_VoigtFit}}
\end{figure*}

In Figure \ref{fig:fix_b_VoigtFit} we present the results of simultaneous Voigt profile fits to \ion{Al}{2} $\lambda$1670, \ion{Fe}{2} $\lambda$1608, \ion{O}{1} $\lambda$1302, \ion{Si}{2} $\lambda$1526 and \ion{Si}{4} $\lambda$1402 absorbers, assuming each absorber has two components. We perform the fits by fixing the doppler parameter $b$ to 5, 10, 15 as well as 30km s$^{-1}$. The column densities and metallicities listed in Table \ref{tab} are concluded from estimates with $b$=5, 10 and 15km s$^{-1}$ , which is the typical $b$ values for metal absorbers seen in high-resolution spectra \citep{Penprase2010}. However, as discussed in Section \ref{sec:dla}, a possible large $b$, e.g., 30km s$^{-1}$, due to possible unresolved sub-components of the absorption profile in low-resolution spectra, would yield a lower column density and thus lower metallicity. 

% \begin{figure*}
%     \centering
%     \begin{subfigure}[b]{0.49\textwidth}
%     \includegraphics[width=\textwidth]{OI+SiII+SiIV_5.fit.pdf}
%     \caption{}
%     \end{subfigure}
%     \begin{subfigure}[b]{0.49\textwidth}
%     \includegraphics[width=\textwidth]{OI+SiII+SiIV_10.fit.pdf}
%     \caption{}
%     \end{subfigure}
%     \begin{subfigure}[b]{0.49\textwidth}
%     \includegraphics[width=\textwidth]{OI+SiII+SiIV_15.fit.pdf}
%     \caption{}
%     \end{subfigure}
%     \begin{subfigure}[b]{0.49\textwidth}
%     \includegraphics[width=\textwidth]{OI+SiII+SiIV_30.fit.pdf}
%     \caption{}
%     \end{subfigure}
%     \caption{Voigt profile fits to the \ion{Al}{2} $\lambda$1670, \ion{Fe}{2} $\lambda$1608, \ion{O}{1} $\lambda$1302, \ion{Si}{2} $\lambda$1526 and \ion{Si}{4}  $\lambda$1402 absorbers, assuming two components for each absorbers and fixing the doppler parameters $b$ to 5, 10, 15 and 30 km s$^{-1}$. Red lines are the best-fit profiles and the blue vertical lines mark the positions of the two components for each absorber. The top row of each panel shows residuals of the best-fit profiles (black) and 3-$\sigma$ uncertainties (red dashed).  The total column density of each absorber, obtained by summing over their two components, is labelled, in units of logN/cm$^{-2}$. }
%     \label{fig:fix_b_VoigtFit}
% \end{figure*}

\bibliography{sample63}{}
\bibliographystyle{aasjournal}

\bigskip

\end{document}